\input harvmac
\input labeldefs.tmp

\newdimen\tableauside\tableauside=1.0ex
\newdimen\tableaurule\tableaurule=0.4pt
\newdimen\tableaustep
\def\phantomhrule#1{\hbox{\vbox to0pt{\hrule height\tableaurule width#1\vss}}}
\def\phantomvrule#1{\vbox{\hbox to0pt{\vrule width\tableaurule height#1\hss}}}
\def\sqr{\vbox{%
  \phantomhrule\tableaustep
  \hbox{\phantomvrule\tableaustep\kern\tableaustep\phantomvrule\tableaustep}%
  \hbox{\vbox{\phantomhrule\tableauside}\kern-\tableaurule}}}
\def\squares#1{\hbox{\count0=#1\noindent\loop\sqr
  \advance\count0 by-1 \ifnum\count0>0\repeat}}
\def\tableau#1{\vcenter{\offinterlineskip
  \tableaustep=\tableauside\advance\tableaustep by-\tableaurule
  \kern\normallineskip\hbox
    {\kern\normallineskip\vbox
      {\gettableau#1 0 }%
     \kern\normallineskip\kern\tableaurule}%
  \kern\normallineskip\kern\tableaurule}}
\def\gettableau#1 {\ifnum#1=0\let\next=\null\else
  \squares{#1}\let\next=\gettableau\fi\next}

\tableauside=1.0ex
\tableaurule=0.4pt
\input epsf
\noblackbox



\def\unlockat{\catcode`\@=11}
\def\lockat{\catcode`\@=12}

\unlockat

\def\newsec#1{\global\advance\secno by1\message{(\the\secno. #1)}
\global\subsecno=0\global\subsubsecno=0\eqnres@t\noindent
{\bf\the\secno. #1}
\writetoca{{\secsym} {#1}}\par\nobreak\medskip\nobreak}
\global\newcount\subsecno \global\subsecno=0
\def\subsec#1{\global\advance\subsecno
by1\message{(\secsym\the\subsecno. #1)}
\ifnum\lastpenalty>9000\else\bigbreak\fi\global\subsubsecno=0
\noindent{\it\secsym\the\subsecno. #1}
\writetoca{\string\quad {\secsym\the\subsecno.} {#1}}
\par\nobreak\medskip\nobreak}
\global\newcount\subsubsecno \global\subsubsecno=0
\def\subsubsec#1{\global\advance\subsubsecno
\message{(\secsym\the\subsecno.\the\subsubsecno. #1)}
\ifnum\lastpenalty>9000\else\bigbreak\fi
\noindent\quad{\secsym\the\subsecno.\the\subsubsecno.}{#1}
\writetoca{\string\qquad{\secsym\the\subsecno.\the\subsubsecno.}{#1}}
\par\nobreak\medskip\nobreak}

\def\subsubseclab#1{\DefWarn#1\xdef
#1{\noexpand\hyperref{}{subsubsection}%
{\secsym\the\subsecno.\the\subsubsecno}%
{\secsym\the\subsecno.\the\subsubsecno}}%
\writedef{#1\leftbracket#1}\wrlabeL{#1=#1}}
\lockat

\def\IL{{\relax{\rm I\kern-.18em L}}}
\def\IH{{\relax{\rm I\kern-.18em H}}}
\def\IR{{\relax{\rm I\kern-.18em R}}}
\def\IE{{\relax{\rm I\kern-.18em E}}}
\def\IP{{\relax{\rm I\kern-.18em P}}}
\def\IC{{\relax\hbox{$\inbar\kern-.3em{\rm C}$}}}
\def\IZ{{\relax\ifmmode\mathchoice
{\hbox{\cmss Z\kern-.4em Z}}{\hbox{\cmss Z\kern-.4em Z}}
{\lower.9pt\hbox{\cmsss Z\kern-.4em Z}}
{\lower1.2pt\hbox{\cmsss Z\kern-.4em Z}}\else{\cmss Z\kern-.4em
Z}\fi}}

\def\CF {{\cal F}}

\def\CO {{\cal O}}

\def\CW {{\cal W}}

\def\CO {{\cal O}}
\def\lu{{\lambda_u}}
\def\lv{{\lambda_v}}

\input epsf
\def\figin{\epsfcheck\figin}\def\figins{\epsfcheck\figins}
\def\epsfcheck{\ifx\epsfbox\UnDeFiNeD
\message{(NO epsf.tex, FIGURES WILL BE IGNORED)}
\gdef\figin##1{\vskip2in}\gdef\figins##1{\hskip.5in}
\else\message{(FIGURES WILL BE INCLUDED)}%
\gdef\figin##1{##1}\gdef\figins##1{##1}\fi}
\def\DefWarn#1{}
\def\figinsert{\goodbreak\midinsert}
\def\ifig#1#2#3{\DefWarn#1\xdef#1{fig.~\the\figno}
\writedef{#1\leftbracket fig.\noexpand~\the\figno}%
\figinsert\figin{\centerline{#3}}\medskip\centerline{\vbox{\baselineskip12pt
\advance\hsize by -1truein\noindent\footnotefont{\bf Fig.~\the\figno } #2}}
\bigskip\endinsert\global\advance\figno by1}

\nref\AAHV{B.~Acharya, M.~Aganagic, K.~Hori and C.~Vafa,
``Orientifolds, Mirror Symmetry and Superpotentials,''
arXiv:hep-th/0202208.}
%
\nref\AKMV{M.~Aganagic, A.~Klemm, M.~Mari\~no and C.~Vafa,
``The Topological Vertex,''
arXiv:hep-th/0305132.}
%
\nref\AMV{M.~Aganagic, M.~Mari\~no and C.~Vafa,
``All Loop Topological String Amplitudes from Chern-Simons Theory,''
Commun.\ Math.\ Phys.\  {\bf 247}, 467 (2004)
[arXiv:hep-th/0206164].}

\nref\AV{
M.~Aganagic and C.~Vafa,
``$G_2$ Manifolds, Mirror Symmetry and Geometric Engineering,''
arXiv:hep-th/0110171.}

\nref\BFM{V.~Bouchard, B.~Florea and M.~Mari\~no,
``Counting Higher Genus Curves with Crosscaps in Calabi-Yau Orientifolds,''
arXiv:hep-th/0405083.}
%
\nref\DF{D.-E.~Diaconescu and B.~Florea, 
``Localization and Gluing of Topological Amplitudes,''
arXiv:hep-th/0309143.}

\nref\DFGi{
D.-E.~Diaconescu, B.~Florea and A.~Grassi, 
``Geometric Transitions and Open String Instantons,''
Adv.\ Theor.\ Math.\ Phys.\  {\bf 6}, 619 (2003)
[arXiv:hep-th/0205234].}
%
\nref\DFGii{
D.-E.~Diaconescu, B.~Florea and A.~Grassi, 
``Geometric transitions, del Pezzo surfaces and open string instantons,''
Adv.\ Theor.\ Math.\ Phys.\  {\bf 6}, 643 (2003)
[arXiv:hep-th/0206163].}
%
\nref\DFM{D.-E.~Diaconescu, B.~Florea and A.~Misra,
``Orientifolds, Unoriented Instantons and Localization,''
JHEP {\bf 0307}, 041 (2003)
[arXiv:hep-th/0305021].}
%
\nref\EK{T. Eguchi and H. Kanno, 
``Topological Strings and Nekrasov's Formulas," 
JHEP {\bf 0312}, 006 (2003) [arXiv: hep-th/0310235]; 
``Geometric Transitions, Chern-Simons Gauge Theory and Veneziano Type
Amplitudes,'' Phys.\ Lett.\ B {\bf 585}, 163 (2004) [arXiv:hep-th/0312234].}
%
\nref\HOMFLY{P. Freyd, D. Yetter, J. Hoste,  W. B. R. Lickorish, K. Millett, and A. Ocneanu,
``A New Polynomial Invariant of Knots and Links,'' Bull. Am. Math. Soc., New Ser. 12, 239-246 (1985).}
\nref\FH{W. Fulton and J. Harris, {\it Representation theory}, 
Springer-Verlag.}
\nref\GV{R.~Gopakumar and C.~Vafa,
``On the Gauge Theory/Geometry Correspondence,''
Adv.\ Theor.\ Math.\ Phys.\  {\bf 3}, 1415 (1999)
[arXiv:hep-th/9811131].}
%
\nref\GViii{R. Gopakumar and C. Vafa, ``M-Theory and Topological Strings. II,''
arXiv:hep-th/9812127.} 
%
\nref\guada{E.~Guadagnini,
``The Universal Link Polynomial,''
Int.\ J.\ Mod.\ Phys.\ A {\bf 7}, 877 (1992).}

%
\nref\HIV{T. Hollowood, A. Iqbal and C. Vafa, 
``Matrix Models, Geometric Engineering and Elliptic Genera," arXiv: hep-th/0310272.}
%
\nref\IKP{A. Iqbal and A.-K. Kashani-Poor, ``Instanton Counting and Chern-Simons Theory," 
Adv.\ Theor.\ Math.\ Phys.\  {\bf 7}, 457 (2004)
[arXiv:hep-th/0212279].}
%
\nref\IKStrip{A.~Iqbal and A.~K.~Kashani-Poor, ``The vertex on a strip,''
arXiv:hep-th/0410174.}

\nref\KL{S. Katz and M. Liu, ``Enumerative Geometry of Stable Maps with Lagrangian Boundary Conditions 
and Multiple Covers of the Disk,'' Adv.\ Theor.\ Math.\ Phys.\  {\bf 5}, 1 (2002)
[arXiv:math.ag/0103074].} 
%
\nref\K{L. H. Kauffman, ``An Invariant of Regular Isotopy,'' 
Trans. Am. Math. Soc. 318, 417 (1990).}
\nref\LLLR{
J.~M.~F.~Labastida, P.~M.~Llatas and A.~V.~Ramallo,
``Knot Operators In Chern-Simons Gauge Theory,''
Nucl.\ Phys.\ B {\bf 348}, 651 (1991).}
%
\nref\LMlink{J.~M.~F.~Labastida and M.~Mari\~no,
``The HOMFLY Polynomial for Torus Links from Chern-Simons Gauge Theory,''
Int.\ J.\ Mod.\ Phys.\ A {\bf 10}, 1045 (1995)
[arXiv:hep-th/9402093].}
%
\nref\LMi{J. M. F. Labastida and M. Mari\~no, ``Polynomial Invariants for Torus Knots and Topological Strings,''
Commun.\ Math.\ Phys.\  {\bf 217}, 423 (2001)
[arXiv:hep-th/0004196].}
%
\nref\LMii{J. M. F. Labastida and M. Mari\~no, ``A New Point of View in the Theory of Knot and Link Invariants,'' 
J. Knot Theory Ramifications {\bf 11}, 173 (2002)
[arXiv:math.qa/0104180].}
%
\nref\LMV{J.~M.~F.~Labastida, M.~Mari\~no and C.~Vafa,
``Knots, Links and Branes at Large $N$,''
JHEP {\bf 0011}, 007 (2000)
[arXiv:hep-th/0010102].}
%
\nref\LP{J.~M.~F.~Labastida and E. P\'erez,
``A Relation Between the Kauffman and the HOMFLY Polynomials for Torus Knots,'' 
J. Math. Phys. {\bf 37}, 2013 (1996) [arXiv:q-alg/9507031].}
%
\nref\LLLZ{J.~Li, C.~C.~Liu, K.~Liu and J.~Zhou, 
``A Mathematical Theory of the Topological Vertex,''
arXiv:math.ag/0408426.}

\nref\L{W. B. R. Lickorish, {\it An Introduction to Knot Theory}, Springer-Verlag, 1998.}
\nref\Li{D. E. Littlewood, {\it The Theory of Group Characters and Matrix Representations of Groups}, Oxford, The Clarendon Press, 1940, 292 p.}
\nref\Mac{I. G. Macdonald, 
{\it Symmetric Functions and Hall Polynomials}, (second edition, 1995), 
Oxford Mathematical Monographs, Oxford Science Publications.}
\nref\MM{M. Mari\~no, ``Chern-Simons Theory and Topological Strings,''
arXiv:hep-th/0406005.}
%
\nref\MV{M.~Mari\~no and C.~Vafa, ``Framed Knots at Large $N$,''
arXiv:hep-th/0108064.}
%
\nref\MNOP{D. Maulik, N. Nekrasov, A. Okounkov and R. Pandharipande, ``Gromov-Witten 
Theory and Donaldson-Thomas Theory, I and II,'' arXiv:math.ag/0312059 and 0406092.}

\nref\ML{H. R. Morton and S. G. Lukac, ``The HOMFLY Polynomial of the 
Decorated Hopf Link," J. Knot Theory Ramifications {\bf 12}, 395 (2003) [arXiv: math.GT/0108011].}
\nref\naculich{S.~G.~Naculich, H.~A.~Riggs and H.~J.~Schnitzer,
``Simple Current Symmetries, Rank Level Duality, and Linear Skein Relations for
Chern-Simons Graphs,'' Nucl.\ Phys.\ B {\bf 394}, 445 (1993)
[arXiv:hep-th/9205082].}

\nref\NV{A.~Neitzke and C.~Vafa, 
``Topological strings and their physical applications,''
arXiv:hep-th/0410178.}

\nref\ORV{A. Okounkov, N. Reshetikhin and C. Vafa, 
``Quantum Calabi-Yau and Classical Crystals," arXiv: hep-th/0309208.}

\nref\OV{H.~Ooguri and C.~Vafa, 
``Knot Invariants and Topological Strings,''
Nucl.\ Phys.\ B {\bf 577}, 419 (2000)
[arXiv:hep-th/9912123].}
%
\nref\rama{P.~Ramadevi, T.~R.~Govindarajan and R.~K.~Kaul, 
``Three-Dimensional Chern-Simons Theory as a Theory of Knots and Links. 3.
Compact Semisimple Group,''
Nucl.\ Phys.\ B {\bf 402}, 548 (1993)
[arXiv:hep-th/9212110].}
\nref\SV{S.~Sinha and C.~Vafa,
``$SO$ and $Sp$ Chern-Simons at Large $N$,''
arXiv:hep-th/0012136.}
%
\nref\T{C. H. Taubes, ``Lagrangians for the Gopakumar-Vafa Conjecture,'' 
Adv. Theor. Math. Phys. {\bf 5},139 (2001) [arXiv:math.DG/0201219].}
\nref\W{E. Witten, ``Quantum Field Theory And The Jones Polynomial,''
Commun.\ Math.\ Phys.\  {\bf 121}, 351 (1989).}
%
\Title{
\vbox{
\baselineskip12pt
\hbox{CERN-PH-TH/2004-229}
\hbox{hep-th/0411227}}}
{\vbox{\vskip 19pt
\vbox{\centerline{Topological Open String Amplitudes on Orientifolds}
}}}
\vskip 15pt
\centerline{Vincent Bouchard\footnote{$^\natural$}
{{{\tt bouchard@maths.ox.ac.uk}}}, Bogdan Florea\footnote{$^\sharp$}
{{{\tt florea@physics.rutgers.edu}}} and Marcos
Mari\~no\footnote{$^\flat$}{{{\tt marcos@mail.cern.ch}; also at Departamento de
Matem\'atica, IST, Lisboa, Portugal.}}}
\bigskip
\medskip
\centerline{$^\natural${\it Mathematical Institute, University of Oxford,}}
\centerline{\it{24-29 St. Giles', Oxford OX1 3LB, England}}
\centerline{$^\sharp${\it Department of Physics and Astronomy,
Rutgers University,}}
\centerline{\it Piscataway, NJ 08855-0849, USA}
\centerline{$^\flat${\it Department of Physics, CERN, Geneva 23, CH-1211, Switzerland}}
\bigskip
\bigskip
\bigskip
\bigskip
\noindent
We study topological open string amplitudes on orientifolds
without fixed planes. We determine the contributions of the untwisted and twisted sectors as well 
as the BPS structure of the amplitudes. We illustrate our general results in 
various examples involving D-branes in toric orientifolds. We perform the computations by using both the 
topological vertex and unoriented localization.
We also present an application of our 
results to the BPS structure of the coloured Kauffman polynomial of knots. 

\vfill
\Date{November 2004}


\writedefs

\newsec{Introduction}

Geometric or large $N$ transitions relating open and closed topological string theories \GV\ have had a deep impact in the 
study of topological strings, and they have led to the first systematic solution of these models 
on noncompact, toric Calabi-Yau threefolds through the topological vertex of \AKMV\ (see \refs{\MM,\NV} for a review).  
The study of topological strings on Calabi-Yau orientifolds was initiated in \SV, where an orientifold of the geometric transition 
of \GV\ relating the deformed and the resolved conifold was studied in detail, and continued in \AAHV\ from the B-model point 
of view. The geometric transition of \GV\ can be extended to more general 
toric geometries \refs{\AV,\DFGi,\DFGii,\AMV}, and in \BFM\ we proposed in fact a general class of large 
$N$ dualities involving orientifolds of non-compact toric Calabi-Yau threefolds. These dualities involve $U(N)$, $SO(N)$ and $Sp(N)$ 
Chern-Simons gauge theories, and they make possible the computation of unoriented string amplitudes. The results obtained 
through large $N$ dualities were also checked in \BFM\ against independent localization computations. 
Moreover, we found a topological vertex prescription to compute these amplitudes directly, extending in this way 
the general formalism of the topological vertex to include the case of orientifolds without fixed planes.

In this paper we continue the study of topological string amplitudes on orientifolds initiated in \BFM. Our main 
goal is to extend the results in \BFM\ to topological open strings on orientifolds without fixed points. In other words, 
we consider orientifolds of non-compact Calabi-Yau threefolds with D-branes. 

An important property of topological string amplitudes is that they have an integrality structure related to the 
counting of BPS states, as it was first realized by Gopakumar and Vafa \GViii\ in the case of closed 
string amplitudes. The integrality structure in the open case was studied in \refs{\OV,\LMV}. As a first step in 
our study of topological string amplitudes on orientifolds without fixed planes we analyze their BPS structure. What we 
find is that the total orientifold amplitude is the sum of an oriented amplitude (the untwisted sector) and 
an unoriented amplitude (the twisted sector) with different integrality properties. We explain how to 
compute the contribution of the twisted sector in the open case. We also spell out in detail the integrality properties 
of the twisted sector contributions. 

This integrality structure provides a strong requirement on topological open string 
amplitudes, and we check it explicitly on various examples involving orientifolds with D-branes. To compute 
these open string amplitudes we use the new vertex rule introduced in \BFM. We also compute the associated Gromov-Witten invariants using 
independent localization techniques developed in \refs{\DFM,\BFM}, and find perfect agreement with the results obtained with the 
vertex.

One of the most interesting applications of the large $N$ duality between open and closed topological strings consists in the determination 
of structural properties of knot and link invariants related to the BPS structure of open topological strings. For example, from the 
results of \refs{\OV,\LMV} one can deduce structure theorems for the coloured HOMFLY polynomial of knots and links. 
The large $N$ duality on orientifolds now involves $SO(N)$ and $Sp(N)$ Chern-Simons theories. Therefore, the BPS structure of the amplitudes 
should lead to the determination of structural properties of a different type of knot and link invariant: the coloured Kauffman polynomial \K. 
Although for arbitrary knots and links we cannot determine in detail the 
structure of the untwisted sector, we are able to derive general structural results for the coloured Kauffman polynomial. 
We test again these predictions on various examples involving torus knots. 

The paper is structured as follows. In section 2, we explore the BPS content of closed and 
open topological string amplitudes on orientifolds and formulate their structural properties. We then compute explicitly 
the amplitudes for various examples in section 3: the $SO/Sp$ framed unknot, the $SO/Sp$ framed Hopf link, and an outer brane in 
$\IP^2$ attached to $\IR \IP^2$. The independent localization computations we provide for all these examples corroborate our methods 
and proposals.
In section 4 we formulate structural properties of the coloured Kauffman polynomial. We discuss our results and propose new avenues of 
research in section 5. Finally, Appendix A contains useful formulae, while in Appendix B we give a full proof 
of the identity that 
was conjectured (and partially proved) in \BFM; this identity shows that the new vertex rule introduced in \BFM\ to compute 
amplitudes on orientifolds agrees with the results of large $N$ $SO/Sp$ transitions. Appendix C lists some results for 
BPS invariants coming from $SO$ Chern-Simons theory.

\newsec{Topological open string amplitudes in orientifolds}

\seclab\structure

\subsec{BPS structure of topological string amplitudes}

One of the most important results of topological string theory is the fact that 
topological string amplitudes have an integrality, or BPS structure, which expresses 
them in terms of numbers of BPS states. Let us briefly review the known results for 
both open and closed strings. 

In the case of topological closed strings on Calabi-Yau threefolds, the BPS structure 
was obtained by Gopakumar and Vafa in \GViii. Let us denote by $F_g(t)$ the topological string 
free energy at genus $g$, where $t$ denotes the set of K\"ahler parameters of the 
Calabi-Yau threefold $X$, 
and let 
\eqn\totalfree{
\CF(t,g_s)=\sum_{g=0}^{\infty} g_s^{2g-2}F_g(t) 
}
be the total free energy. Then, one has the following structure result:
\eqn\gv{
\CF(t,g_s)=\sum_{d=1}^{\infty}\sum_{g=0}^{\infty}\sum_{\beta} 
{1\over d} { n^g_\beta \over (q^{d\over 2} - q^{-{d\over 2}})^{2-2g}}
e^{-d\beta\cdot t}.
} 
where $q=e^{i g_s}$, the sum over $\beta$ is over two-homology classes in $X$, and $n^g_\beta$ (the so-called Gopakumar-Vafa 
invariants) are {\it integers}. The factor $(q^{d\over 2} - q^{-{d\over 2}})^{2g}$ comes from computing 
a signed trace over the space of differential forms on a Riemann surface of genus $g$, while the factor 
$(q^{d\over 2} - q^{-{d\over 2}})^{-2}$ comes from a Schwinger computation \GViii. 

For open string amplitudes, the structure of the amplitudes was found in \refs{\OV,\LMV} and is much more delicate.  
To define an open string amplitude we have to specify boundary conditions through a set of submanifolds of $X$, $S_1, 
\cdots, S_L$. To each of these submanifolds we associate a source $V_{\ell}$, $\ell=1, \cdots, L$, which is 
a $U(M)$ matrix. The total partition function is given by
\eqn\partf{
Z(V_1, \cdots, V_L)=\sum_{R_1, \cdots, R_L}Z_{(R_1, \cdots, R_L)}
\prod_{\alpha=1}^L {\rm Tr}_{R_{\alpha}} V_{\alpha},}
where $R_{\alpha}$ denote representations of $U(M)$ and we are considering
the limit $M\rightarrow \infty$. 
The amplitudes $Z_{(R_1, \cdots, R_L)}$ can be computed in the noncompact, toric case by using 
the topological vertex \AKMV. According to the correspondence proposed in \OV, they are given 
in some cases by invariants of links whose components are coloured by representations $R_1, \cdots, R_L$. 
The free energy is defined as usual by 
\eqn\freeopen{
\CF(V_1, \cdots, V_L)= - \log \, Z(V_1, \cdots, V_L)
}
and is understood as a series in traces of $V$ in different representations.   
We define the generating function
$f_{(R_1, \cdots, R_L)} (q, \lambda)$ through the
following equation:
\eqn\conj{
{\cal F}(V)= - \sum_{n=1}^\infty \sum_{R_1, \cdots, R_L}
{1\over n} f_{(R_1, \cdots, R_L)} (q^n, {\rm e}^{-n t})
\prod_{\alpha=1}^L {\rm Tr}_{R_\alpha}V_{\alpha}^n
 }
The main result of \LMV\ is that 
$f_{(R_1, \cdots, R_L)} (q, e^{-t})$ is given by:
\eqn\fr{
\eqalign{
&f_{(R_1, \cdots, R_L)}(q, e^{-t})=\cr &
(q^{1\over 2}-q^{-{1 \over 2}})^{L-2}
\sum_{g\ge 0} \sum_{\beta}
\sum_{R'_1, R_1'' \cdots, R'_L, R_L''}  \prod_{\alpha=1}^L
c_{R_{\alpha}\,R_{\alpha}'\,R_{\alpha}''}S_{R_{\alpha}'}(q)
N_{(R_1'', \cdots, R_L''),g,\beta}
 (q^{1\over 2}-q^{-{1\over 2}})^{2g} {\rm e}^{-\beta\cdot t}.\cr}}
In this formula $R_{\alpha},R_{\alpha}',R_{\alpha}''$ label
representations of the symmetric group
$S_\ell$, which can be labeled
by a Young tableau with a total of $\ell$ boxes.
$c_{R\,R'\,R''}$ are the Clebsch-Gordon coefficients
of the symmetric group, and the monomials $S_R (q)$ are defined as
follows. If $R$ is a hook representation
\eqn\hook{
\tableau{6 1 1 1 1}}
with $\ell$ boxes in total, and with $\ell-d$ boxes in the first row, then
\eqn\expsr{
S_R (q)=(-1)^d q^{ -{\ell -1 \over 2}+d} ,}
and it is zero otherwise.
Finally, $N_{(R_1, \cdots, R_L),g,\beta}$ are {\it integers}
associated to open string amplitudes. They compute the net number
of BPS domain walls of charge $\beta$ and spin $g$ transforming
in the representations $R_{\alpha}$ of $U(M)$, where we are using
the fact that representations of $U(M)$ can also be labeled by Young
tableaux. It is also useful to introduce a generating functional for these
degeneracies as in \LMV:
\eqn\tildefrlinks{
{\widehat f}_{(R_1, \cdots, R_L)}(q, e^{-t})=
\sum_{g \ge 0}\sum_\beta N_{(R_1, \cdots, R_L),g,\beta}
(q^{{1\over 2}}-q^{-{1 \over 2}})^{2g+L-2}{\rm e}^{-\beta\cdot t}.}
We then have the relation:
\eqn\relafslinks{
 f_{(R_1, \cdots, R_L)}(q,e^{-t})=
\sum_{R_1', \cdots, R'_L}\prod_{\alpha=1}^L 
M_{R_{\alpha} R'_{\alpha}}(q)
{\widehat f}_{(R_1, \cdots, R_L)}(q,e^{-t} ),}
where the matrix $M_{R R'}(q)$ is given by
$$
M_{R R'}(q)= \sum_{R''} c_{R\,R'\,R''}S_{R''}(q)
$$
and it is symmetric and invertible \LMV. 
The $f_{(R_1, \cdots, R_L)}$ introduced in \conj\ can be extracted
from $Z_{(R_1, \cdots, R_L)}$ through a procedure spelled out in detail in
\refs{\LMi,\LMii,\LMV}. One has, for example,
\eqn\exm{
f_{\tableau{1} \tableau{1}}=
Z_{\tableau{1} \tableau{1}}-Z_{\tableau{1} 0} Z_{0 \tableau{1}},}
where $0$ denotes the trivial representation. 
As it was emphasized in \refs{\LMi,\LMii,\LMV}, this structure result has interesting 
consequences for knot theory, since it implies a series of integrality results for 
knot and link invariants. We will come back to this issue in section \kauffman.

\subsec{BPS structure of topological strings on orientifolds}

We want to understand now the corresponding BPS structure of closed and 
open topological string 
amplitudes on orientifolds without fixed points, like the ones considered in \refs{\SV,\AAHV}. 
In \BFM\ the closed case was studied in detail, in the noncompact case, by using large $N$ 
transitions and the topological vertex. Let us denote by $X/I$ the orientifold obtained by 
an involution on $X$. The total free energy has in this case the 
structure
\eqn\strucf{
\CF(X/I, g_s)= {1\over 2}\CF(X, g_s) + \CF(X/I, g_s)_{\rm unor},
}
where $g_s$ is the string coupling constant. In the r.h.s. of this equation, 
the first summand is the contribution of the untwisted sector,
and it involves the free energy $\CF(X, g_s)$ of the covering $X$ of $X/I$, 
after suitably identifying
the K\"ahler classes in the way prescribed by the involution $I$. This piece of 
the free energy has an expansion identical to \gv, but due to the factor $1/2$ it 
involves half-integers instead of integers. The second 
summand, that we call the unoriented part $\CF( X/I, g_s)_{\rm unor}$, is the contribution of the twisted sector, 
and involves the counting of holomorphic maps from
closed non-orientable Riemann surfaces to the orientifold $X/I$.
The Euler characteristic of a closed Riemann surface of genus $g$ and $c$ crosscaps
is $\chi = -2g+2-c$ where $c$ is the number of crosscaps. We then have
\eqn\strucun{
\CF(X/I, g_s)_{\rm unor}= \CF(X/I, g_s)_{\rm unor}^{c=1}+ \CF(X/I, g_s)_{\rm unor}^{c=2},}
which corresponds to the contributions of one and two crosscaps. Following
the arguments in \GViii\ we predict the following structure
\eqn\strucuntwo{\eqalign{
\CF(X/I, g_s)_{\rm unor}^{c=1}=&
\pm \sum_{d \, \, {\rm odd}}\sum_{g=0}^{\infty}\sum_{\beta }  n^{g,c=1}_\beta
{1\over d}(q^{d\over 2} - q^{-{d\over 2}})^{2g-1}e^{-d \beta\cdot t},\cr
\CF(X/I, g_s)_{\rm unor}^{c=2}=&
\sum_{d \, \, {\rm odd}} \sum_{g=0}^{\infty}\sum_{\beta }
n^{g,c=2}_\beta {1\over d}(q^{d\over 2} - q^{-{d\over 2}})^{2g}e^{-d \beta\cdot t},\cr}
}
where $n^{g,c}_\beta$ are integers. The $\pm$ sign in the $c=1$ free energy is due to
the two different choices for the sign of the crosscaps, and the restriction to 
$d$ odd comes, in the case of $c=1$, from the geometric absence of even 
multicoverings. In the $c=2$ case this was concluded from examination of 
different examples. The structure results in \strucf, \strucun\ and \strucuntwo\ were tested in \BFM\ through detailed 
computations in noncompact geometries.  

We now address the generalization to open string amplitudes in orientifolds. We first consider 
for simplicity the case of a single boundary condition in the orientifold $X/I$ associated to a topological 
D-brane wrapping a submanifold 
$S$. As in the closed string case, the total open string amplitude will have a contribution from 
untwisted sectors, and a contribution from twisted sectors. We will then write
\eqn\freen{
\CF(V)={1\over 2} \CF_{\rm or} (V) + \CF_{\rm unor}(V),}
The contribution from the untwisted sector, $F_{\rm or} (V)$, 
involves the covering geometry, which will be given by $X$, the submanifold $S$, and its image under the involution 
$I(S)$. In other words, the covering amplitude will involve now {\it two} different sets of D-branes, in general. The 
covering geometry with two sets of branes has the total partition function
\eqn\covam{
Z_{\rm cov}(V_1, V_2)=\sum_{R_1, R_2} {\cal C}_{R_1 R_2} {\rm Tr}_{R_1}V_1 \, {\rm Tr}_{R_2} V_2,
}
where $V_1$, $V_2$ are the sources corresponding to $S$ and $I(S)$ and represent open string 
moduli. Since the two D-branes in $S$ and $I(S)$ are related by an involution, the two-brane 
amplitude in \covam\ is symmetric under their exchange, {\it i.e.} we have 
\eqn\sime{
{\cal C}_{R_1 R_2}={\cal C}_{R_2 R_1}.
}
In order to obtain ${\cal F}_{\rm or}(V)$ we have to make the identification of {\it both} closed and open 
string moduli under the involution $I$. This means identifying the K\"ahler parameters that appear in 
${\cal C}_{R_1 R_2}$ (the closed background) but also setting $V_1=V_2=V$ (the open background). We then find
\eqn\zcov{
Z_{\rm or}(V)=\sum_R Z_R^{\rm or} {\rm Tr}_R V}
where
\eqn\zror{
Z_R^{\rm or}=\sum_{R_1, R_2} N^R_{R_1 R_2} {\cal C}_{R_1 R_2} =\sum_{R'} {\cal C}_{R/R' R'}.}
Here we have used that
\eqn\tensorp{
{\rm Tr}_{R_1}V \, {\rm Tr}_{R_2} V =\sum_{R}N^R_{R_1 R_2} \, {\rm Tr}_{R} V}
and $N^R_{R_1 R_2}$ are tensor product coefficients. In \zror\ we also used these 
coefficients to define skew coefficients with labels $R/R'$, as in \skewschur. 
If we denote ${\cal C}_R \equiv {\cal C}_{R\cdot}$, we 
have for example
\eqn\listz{
Z_{\tableau{1}}^{\rm or} =2\, {\cal C}_{\tableau{1}},\quad
Z_{\tableau{2}}^{\rm or} = 2 \, {\cal C}_{\tableau{2}} +
{\cal C}_{\tableau{1} \tableau{1}},\quad
Z_{\tableau{1 1}}^{\rm or} =2{\cal C}_{\tableau{1 1}}+
{\cal C}_{\tableau{1} \tableau{1}}.}
It turns out that the quantities $Z^{\rm or}_R$ defined in this way 
have the integrality properties of a one-brane 
amplitude, as it should. One finds, for example,
\eqn\listf{
{\widehat f}_{\tableau{1}}^{\rm or} =2 \widehat f^{\rm cov}_{\tableau{1} \cdot},\quad
{\widehat f}_{\tableau{2}}^{\rm or} = 2 \widehat f^{\rm cov}_{\tableau{2} \cdot}-{1\over
q^{1\over 2} -q^{-{1\over 2}}} \widehat f^{\rm cov}_{\tableau{1} \tableau{1}}, \quad
{\widehat f}_{\tableau{1 1}}^{\rm or} =2 \widehat f^{\rm cov}_{\tableau{1 1} \cdot}-{1\over
q^{1\over 2} -q^{-{1\over 2}}} \widehat f^{\rm cov}_{\tableau{1} \tableau{1}}. \quad}
In these equations, the superscript ``${\rm cov}$'' refers to quantities computed from the two-brane 
amplitude ${\cal C}_{R_1 R_2}$ according to the general rules for open string amplitudes in the usual, oriented 
case explained above. One can easily verify from the integrality properties of $\widehat f_{R_1 R_2}$ as a
2-brane amplitude that indeed
${\widehat f}_R^{\rm or}$ has the integrality properties of a one-brane
amplitude. In fact, using the identity
\eqn\idef{
\sum_{R', R_1', R_2'} M_{R R'}^{-1} N^{R'}_{R_1' R_2'} M_{R'_1 R_1} M_{R'_2 R_2}= 
{1\over q^{-{1\over 2}}- q^{1\over 2}} N^R_{R_1 R_2}
}
we can write
\eqn\covint{
{\widehat f}_R^{\rm or}=\sum_{R_1 R_2} N^R_{R_1 R_2} {\widehat f}^{\rm cov}_{R_1 R_2},}
where we put ${\widehat f}_{R 0}\equiv (q^{-{1\over 2}}- q^{1\over 2}) {\widehat f}_R$. 

We would like to determine now the structural properties of $\CF_{\rm unor}(V)$. This is indeed very easy. 
The analysis of \LMV\ to determine the structural properties of $F(V)$ in the usual oriented case was based 
on an analysis of the Hilbert space associated to an oriented Riemann surface $\Sigma_{g,\ell}$ with $\ell$ 
holes ending on $S$ and in the relative homology class $\beta \in H_2(X,S)$. 
The relevant Hilbert space turns out to be
\eqn\hilbert{
{\rm Sym}\bigl(F^{\otimes \ell} \otimes H^*(J_{g,\ell}) \otimes
H^*({\cal M}_{g,\ell, \beta})\bigr)
}
where $J_{g,\ell}={\bf T}^{2g+\ell-1}$ is the Jacobian of $\Sigma_{g,\ell}$, $F$ is a 
copy of the fundamental representation of the gauge group, ${\cal M}_{g,\ell, \beta}$ is the 
moduli space of geometric deformations of the Riemann surface inside the Calabi-Yau manifold, and ${\rm Sym}$ 
means that we take the completely symmetric piece with respect to permutations of the 
$\ell$ holes. Since the bulk of the Riemann surface is not relevant for the action of 
the permutation group, we can
factor out the cohomology of the Jacobian ${\bf T}^{2g}$. The projection onto the
symmetric piece can easily be done using the Clebsch-Gordon
coefficients $c_{R\,R'\,R''}$ of the permutation group $S_{\ell}$ \FH, 
and one finds 
\eqn\decomhilbert{
\sum_{R\,R'\,R''} c_{R\,R' \, R''} {\bf
S}_R(F^{\otimes \ell})\otimes{\bf S}_{R'}(H^*(({\bf
S}^1)^{\ell-1})) \otimes {\bf S}_{R''}(H^*({\cal M}_{g,\ell,\beta}))}
where ${\bf S}_R$ is the Schur functor that projects onto
the corresponding subspace. The space ${\bf S}_R(F^{\otimes
\ell})$ is nothing but the vector space underlying the irreducible
representation $R$ of $U(M)$. ${\bf S}_{R'}(H^*(({\bf
S}^1)^{\ell-1}))$ gives the hook Young tableau, and the Euler
characteristic of ${\bf S}_{R''}(H^*({\cal M}_{g,\ell, \beta}))$ is
the integer invariant $N_{R'',g,\beta}$. Therefore, the above
decomposition corresponds very precisely to \fr\ (here we are considering 
for simplicity the one-brane case). 

In the case of an {\it unoriented} Riemann surface, the above argument goes through, 
with the only difference that now the Jacobian is $J_{g,c,\ell}={\bf T}^{2g-1+\ell +c}$, 
where $c=1,2$ denotes the number of crosscaps. Therefore, the analysis of the cohomology associated 
to the boundary is the same. We then conclude that
\eqn\freeunor{
\CF_{\rm unor}(V)= - \sum_R \sum_{d \, {\rm odd}}{1\over d}f_R^{\rm unor} (q^d, e^{-d t})
{\rm Tr}_R V^d,}
and using again \relafslinks\ one can obtain new functions
\eqn\unorinv{
\widehat f^{\rm unor}_R=\sum_R' M^{-1}_{R R'} f_{R'}^{\rm unor}
}
with 
contributions from one and two crosscaps: 
\eqn\ccont{
{\widehat f}^{\rm unor}_R={\widehat f}^{c=1}_R+ (q^{1\over 2} -q ^{-{1\over 2}}){\widehat f}^{c=2}_R,}
and we finally have
\eqn\cs{
\widehat f_R^c (q, e^{-t})=\sum_{g,\beta} N^c_{R,g,\beta} (q^{1\over 2} -q ^{-{1\over 2}})^{2g}
e^{-\beta\cdot t}.
}
Each crosscap contributes then an extra factor of $q^{1\over 2} -q^{-{1\over 2}}$, as in the closed case.

In real life, what one computes is the total amplitude in the l.h.s. of \freen, in terms of 
\eqn\totalf{
{\cal F}(V)=- \log Z(V)=- \log \Bigr( \sum_R Z_R {\rm Tr}_R \, V \Bigl),}
and one wants to find the unoriented contribution to the amplitude after subtracting the oriented 
contribution. The above formulae give a precise prescription to compute $f_R^{\rm 
unor}$. The results one finds, up to three boxes, 
are the following:
\eqn\threeres{
\eqalign{
f^{\rm unor}_{\tableau{1}}=&Z_{\tableau{1}}-{\cal C}_{\tableau{1}}, \cr
f^{\rm unor}_{\tableau{2}}=&Z_{\tableau{2}}-{1\over 2}Z_{\tableau{1}}^2 -{\cal C}_{\tableau{2}} +{1\over 2}
{\cal C}_{\tableau{1}}^2 
-{1\over 2}f^{\rm cov}_{\tableau{1} \tableau{1}}, \cr 
f^{\rm unor}_{\tableau{1 1}}=&Z_{\tableau{1 1}}- {1\over 2}Z_{\tableau{1}}^2 -{\cal C}_{\tableau{1 1}} +
{1\over 2}{\cal C}_{\tableau{1}}^2-{1\over 2}f^{\rm cov}_{\tableau{1} \tableau{1}}, \cr
f^{\rm unor}_{\tableau{3}}=&Z_{\tableau{3}}-Z_{\tableau{2}} Z_{\tableau{1}} +{1 \over 3} Z^3_{\tableau{1}}-{\cal C}_{\tableau{3}}
+{\cal C}_{\tableau{2}} {\cal C}_{\tableau{1}} -{1 \over 3} {\cal C}^3_{\tableau{1}} \cr
&-{1\over 3} f^{\rm unor}_{\tableau{1}}(q^3, Q^3)-f^{\rm cov}_{\tableau{2} \tableau{1}},\cr 
}}
and
\eqn\threeresi{\eqalign{
f^{\rm unor}_{\tableau{2 1}}=&Z_{\tableau{2 1}}-Z_{\tableau{2}} Z_{\tableau{1}} -Z_{\tableau{1 1}} Z_{\tableau{1}} 
+{2 \over 3} Z^3_{\tableau{1}}-{\cal C}_{\tableau{2 1}}+{\cal C}_{\tableau{2}} {\cal C}_{\tableau{1}} +{\cal C}_{\tableau{1 1}} 
{\cal C}_{\tableau{1}} +{2 \over 3} {\cal C}^3_{\tableau{1}}\cr &+{1\over 3} f_{\tableau{1}}^{\rm unor} (q^3,Q^3) - {1\over 2}
(f^{\rm cov}_{\tableau{2} \tableau{1} } +f^{\rm cov}_{\tableau{1 1} \tableau{1}}),\cr
f^{\rm unor}_{\tableau{1 1 1}}=&Z_{\tableau{1 1 1}}-Z_{\tableau{1 1}} Z_{\tableau{1}} +{1 \over 3} Z^3_{\tableau{1}}
-{\cal C}_{\tableau{1 1 1}}-{\cal C}_{\tableau{1 1}} {\cal C}_{\tableau{1}} +{1 \over 3} {\cal C}^3_{\tableau{1}}\cr
&-{1\over 3} f^{\rm unor}_{\tableau{1}}(q^3, Q^3)-f^{\rm cov}_{\tableau{1 1} \tableau{1} }.\cr}
}

The above considerations are easily extended to the case in which we have $L$ sets of 
D-branes in the orientifold geometry. The covering 
amplitude involves now $2L$ D-branes, and reads
\eqn\covaL{
Z_{\rm cov}=\sum_{R_1, S_1, \cdots, R_L, S_L} {\cal C}_{R_1 S_1 \cdots R_L S_L} {\rm Tr}_{R_1}V_1{\rm Tr}_{S_1}W_1  
\cdots {\rm Tr}_{R_L}V_1{\rm Tr}_{S_L} 
W_L.
}
The oriented amplitude is obtained by identifying the moduli in pairs under $I$, and is given by
\eqn\zrorL{
Z_{Q_1 \cdots Q_L}^{\rm or}=\sum_{R_i, S_i} N^{Q_1}_{R_1 S_1} \cdots N^{Q_L}_{R_L S_L} {\cal C}_{R_1 S_1 \cdots R_L S_L} .}
The equations \freen, \freeunor\ and \cs\ generalize in an obvious way, but now we have
\eqn\linkc{
\widehat f_{(R_1 \cdots R_L)}^c 
(q, e^{-t})=\sum_{g,\beta} N^c_{(R_1, \cdots, R_L),g,\beta} (q^{1\over 2} -q ^{-{1\over 2}})^{2g+L-1}
e^{-\beta\cdot t},
}
where the extra $L-1$ factors of $q^{1\over 2} -q ^{-{1\over 2}}$ have the same origin as in \tildefrlinks.

\newsec{Examples of open string amplitudes} 

\seclab\examples

In this section we study in detail some examples and verify the above formulae for the 
unoriented part of the free energy. In order to do that, we have to compute the total amplitudes 
$Z_R$ in orientifold geometries. These amplitudes can be obtained in three ways: by using the unoriented 
localization methods of \refs{\DFM,\BFM}, by using mirror symmetry \AAHV, and by using Chern-Simons theory 
and the topological vertex. For the examples of open string amplitudes studied in this section 
we will use the topological vertex of \AKMV, which can be adapted to the orientifold case \BFM, and 
also localization. We first summarize very briefly the results of \BFM\ on the topological vertex on orientifolds, 
and then we study in detail three examples. Finally we check some of the topological vertex results with 
unoriented localization.   

\vfill\eject

\subsec{The topological vertex on orientifolds}

\ifig\vertexfig{Toric diagram for the quotient $X/I$ of a local, toric Calabi-Yau manifold
with a single ${\IR\IP}^2$.}
{\epsfxsize2in\epsfbox{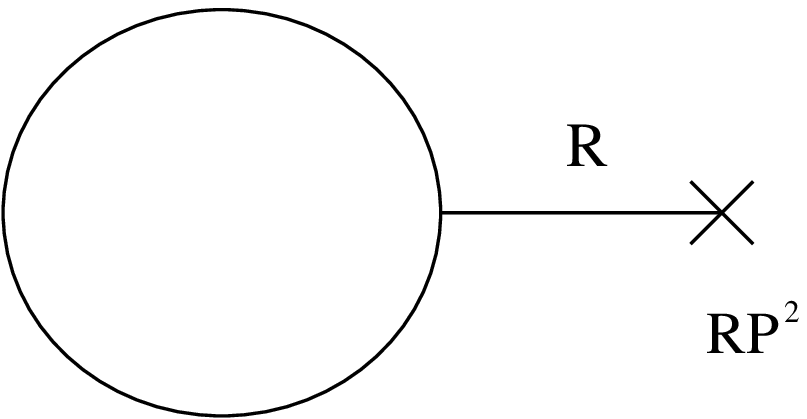}}
Let us consider a quotient $X/I$ of a local, toric Calabi-Yau manifold $X$ by an involution $I$
without fixed points which can be represented as in \vertexfig. We have a
bulk geometry, represented by the blob, attached to an ${\IR\IP}^2$ through an
edge with representation $S$. Let us denote by
${\cal O}_{S}$ the amplitude for the blob with the external leg. In \BFM\
we proposed the following formula for the total partition function:
\eqn\vertextotal{
Z=\sum_{S=S^T} \CO_S  Q^{\ell(S)/2}
(-1)^{{1\over 2}(\ell(S)\mp r(S))}
}
where the sum is over all self-conjugate
representations $S$. Here $r(S)$ denotes the rank of $S$, $Q$ is the exponentiated
K\"ahler parameter corresponding to the ${\IR\IP}^2$, and the $\mp$ sign is correlated with the
choice of $\pm$ sign for the crosscaps. In \BFM\ the above prescription was used to compute closed 
string amplitudes, and we conjectured the following identity:
\eqn\magic{
{1\over S_{00}^{SO(N)/Sp(N)}}\sum_{R=R^T}   C_{R_1 R_2^T R} Q^{\ell(R)/2}
(-1)^{{1\over 2}(\ell(R)\mp r(R))} = q^{-{\kappa_{R_2}\over 2}}
Q^{{1\over 2}(\ell(R_1) + \ell (R_2))}
{\cal W}^{SO(N)/Sp(N)}_{R_1 R_2},}
 where $C_{R_2^T R R_1}$ is the topological vertex of \AKMV, 
${\cal W}^{SO(N)/Sp(N)}_{R_1 R_2}$ is the $SO$/$Sp$ Chern-Simons expectation value 
of the Hopf link with linking number $+1$ (after setting the Chern-Simons variable $\lambda$ 
defined in \lam\ to be $\lambda=Q^{-1}$), 
and $S_{00}^{SO(N)/Sp(N)}$ is the partition function of $SO$/$Sp$ Chern-Simons theory on ${\bf S}^3$. 
We refer the reader to \refs{\MM,\BFM} for explicit formulae for the Hopf 
link invariants. The identity \magic\ was used in \BFM\ to show that the results obtained with \vertextotal\ coincide with those 
that are obtained from large $N$ transitions involving $SO$/$Sp$ gauge groups. In the 
examples that follow we will use \vertextotal\ to compute open string amplitudes on orientifolds, 
making use as well of the identity \magic. We provide a proof of \magic\ in Appendix B.  
\vskip1in

\subsec{The $SO/Sp$ framed unknot}

\ifig\dorienti{A D-brane in an outer leg of the orientifold of 
the resolved conifold.}
{\epsfxsize 2.0truein\epsfbox{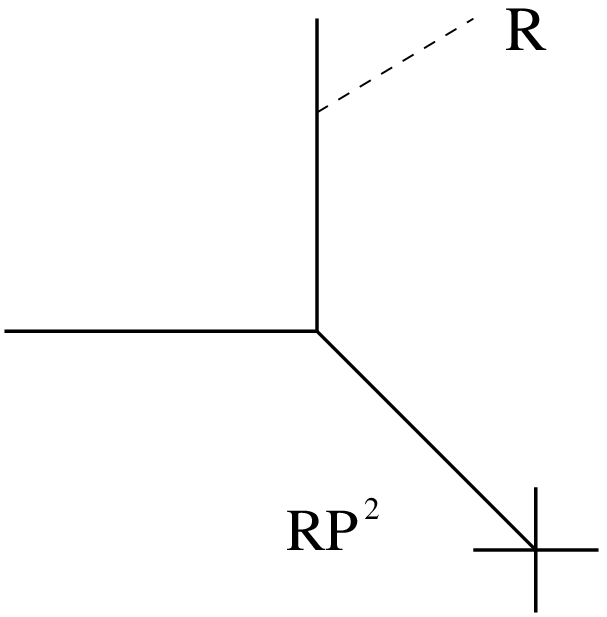}} 

We start with the simplest nontrivial Calabi-Yau orientifold, namely the orientifold of the 
resolved conifold. The resolved conifold $X$ is a noncompact Calabi-Yau
threefold which admits a toric description given by the following toric data:
\eqn\toricres{
\matrix{ &  X_1 & X_2 & X_3 & X_4  \cr
{\bf C}^* & 1 & 1 & -1 & -1 }}
Therefore, it is defined as the space obtained from
\eqn\rescon{
|X_1|^2 + |X_2|^2 -|X_3|^2 - |X_4|^2 =t
}
after quotienting by the $U(1)$ action specified by the charges in \toricres. 
The involution 
\eqn\orres{
I: (X_1, X_2, X_3, X_4) \rightarrow ({\overline X}_2, -{\overline X}_1,
{\overline X}_4, -{\overline X}_3).
}
leads to an orientifold model whose target $X/I$ contains a single ${\IR\IP}^2$ obtained from the
quotient of the ${\IP}^1$ of $X$ by $I$. This geometry was first considered in \refs{\SV,\AAHV} and further studied in 
\BFM. 

Let us now put a D-brane in an {\it outer} leg of the orientifold geometry. In the oriented case, the open string 
amplitude labelled by $R$ is computed by the Chern-Simons invariant of 
the framed unknot with gauge group $U(N)$ (see for example \refs{\MM,\MV}). 
We want to study now the unoriented case. In order to extract the unoriented 
string amplitudes, we have to compute both the total amplitudes $Z_R$ and the 
covering amplitudes ${\cal C}_{R_1 R_2}$. Let us start analyzing the covering amplitude. 
 
\ifig\dcover{The covering configuration contains two D-branes.}
{\epsfxsize 3.5truein\epsfbox{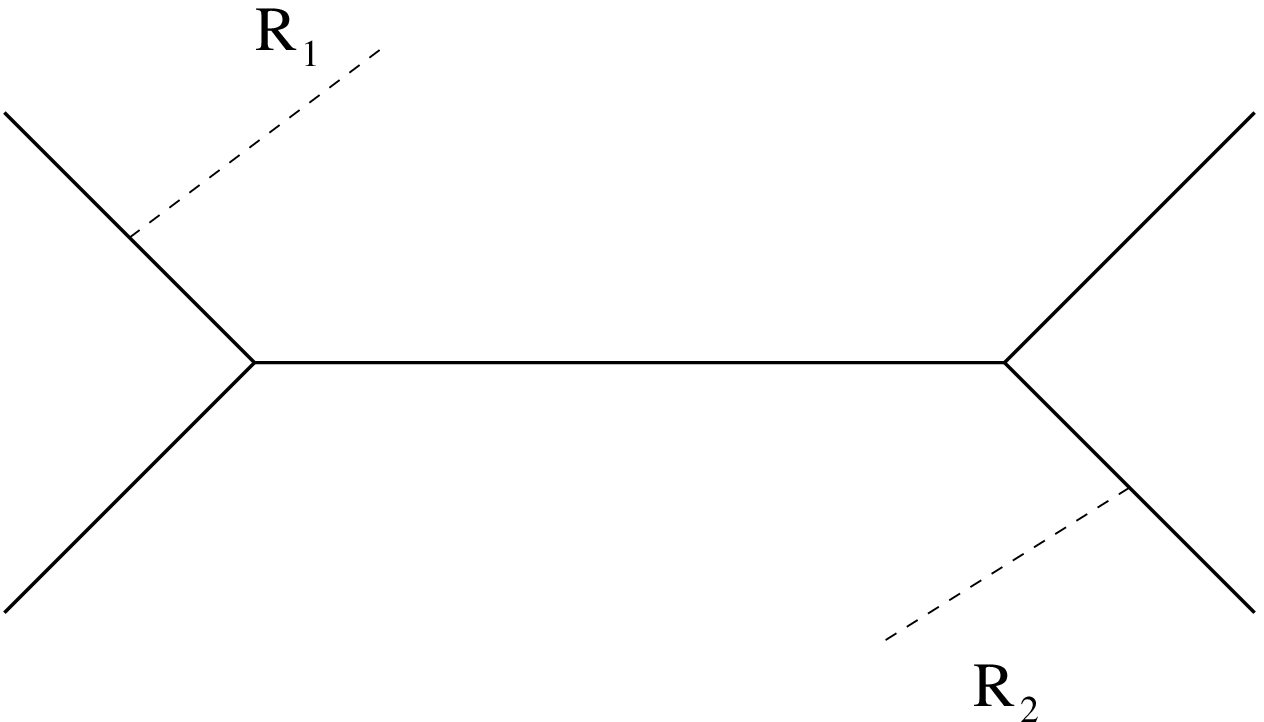}} 

The covering geometry involves both the original D-brane and its image under the 
involution $I$, and a simple analysis shows that we have to consider {\it two} D-branes in opposite legs as 
depicted in \dcover. The amplitude for this two-brane configuration can be 
computed by using the topological vertex of \AKMV\ (see Appendix A for a list of 
useful formulae and properties of the vertex). A simple application of the rules in \AKMV\ gives
\eqn\cfirst{
{\cal C}_{R_1 R_2}={1 \over Z_{{\IP}^1}}\sum_R C_{R_1 R \cdot}C_{R^T \cdot R_2} (-Q)^{\ell(R)}={1 \over Z_{{\IP}^1}}
\sum_{R} W_{R_1 R^T}
W_{R R_2} (-Q)^{\ell(R)}
}
where $Q=e^{-t}$, $Z_{\IP^1}$ is the partition function of the resolved conifold
\eqn\zconif{
Z_{\IP^1} = \prod_{k=1}^{\infty} (1-Q q^k)^k,
}
and the quantities $W_{R_1 R_2}$ are defined in \hopfschur. The above quotient of series 
can be computed in a closed way by using the techniques of \refs{\IKP,\EK}, and in fact one obtains two 
equivalent expressions.
The first expression is
\eqn\csecond{
{\cal C}_{R_1 R_2} =W_{R_1} W_{R_2} \prod_k (1- q^k Q)^{C_k (R_1, R_2)}
}
where the coefficients $C_k (R_1, R_2)$ are given by \expcoeff\ or \coeffs. Notice that \csecond\ is a 
Laurent polynomial in $q^{\pm {1 \over 2}}$ and a
{\it polynomial} in $Q$. There is, however, a second expression for ${\cal C}_{R_1 R_2}$ which involves skew quantum dimensions.
The derivation uses the representation of the vertex in terms of skew Schur functions given in \topvertexschur.
Define first:
\eqn\renquan{
{\cal W}_R=  ({\rm dim}_q^{U(N)} R)(\lambda=Q^{-1}),}
where $\lambda$ is again the Chern-Simons variable \lam, and the quantum dimension is defined in \qdgeneral. 
Then we have, after using \identitythree,
\eqn\cthree{
{\cal C}_{R_1 R_2}= q^{{ \kappa_{R_1} + \kappa_{R_2} \over 2}}Q^{\ell (R_1)+ \ell(R_2)\over 2}
\sum_R (-1)^{\ell (R)} {\cal W}_{R_1^T/R} {\cal W}_{R_2^T/R^T}.}

We now compute the total amplitude for the configuration 
depicted in \dorienti. To do this we can use \vertextotal, 
where ${\cal O}_S$ is now an open string amplitude given by $C_{\cdot RS}$. One finds
\eqn\fone{\eqalign{
Z_R=&{1 \over Z_{X/I}} \sum_{R'=R'^T}C_{\cdot R R'} Q^{\ell (R')/2}(-1)^{{1\over 2}(\ell(R')\mp r(R'))}\cr
=&q^{\kappa_R\over 2} {\cal W}_{R^T}^{SO/Sp},\cr}}
where we have used the formula \magic\ to express the amplitude in terms of $SO/Sp$ quantum dimensions. 
We then see that the {\it total} brane amplitude in \dorienti\ is given by the Chern-Simons invariant of 
an unknot for gauge groups $SO/Sp$. To obtain the unoriented piece of this amplitude, we have 
to subtract the covering contribution, which involves a nontrivial combination of quantum dimensions for $U(N)$. 
For a {\it framed} D-brane one should simply change
\eqn\framingch{
\eqalign{
Z_R &\rightarrow (-1)^{\ell (R)p } q^{p\kappa_R\over 2}Z_R, \cr
{\cal C}_{R_1 R_2} &\rightarrow (-1)^{(\ell (R_1)+ \ell(R_2))p } q^{p{\kappa_{R_1}
+\kappa_{R_2}\over 2}}{\cal C}_{R_1 R_2},\cr}}
since in the covering configuration one has to put
the same framing in both legs, by symmetry.

We can now compute $f^{\rm unor}_R$ by using the results of 
the previous section. We will present explicit results only up to 
three boxes. The first thing one finds is that $f^{c=2}_R$ vanishes at this 
order in $R$. For $f^{c=1}_R$ one finds (we present here the results for $SO(N)$; for $Sp(N)$ one only has to change the overall sign of the $c=1$ contributions):
$$
\eqalign{
\widehat f^{c=1}_{\tableau{1}}=& (-1)^p Q^{1/2},\cr
\widehat f^{c=1}_{\tableau{2}}=&{q^{1 - p}\,\left( 1 - q^{p} - q^{1 + p} + 
    q^{1 + 2\,p} \right) \,Q^{1/2}\,\left( -1 + Q \right)\over (q -1)^2 (q +1)},\cr
\widehat f^{c=1}_{\tableau{1 1}}=&q^{-p} {\left( 1 - q^{1 + p} - q^{2 + p} + 
    q^{3 + 2\,p} \right) \,Q^{1/2}\,\left( -1 + Q \right)\over (q -1)^2 (q +1)},\cr
}
$$
and
$$
\eqalign{
\widehat f^{c=1}_{\tableau{3}}=&{ (-1)^p q^{2-3p} (-1+q^p) (-1+q^{1+p}) Q^{1/2} (-1+Q) \over(-1+q)^4 (1+q)^2 (1+q^2+q^4)} \cr
&\times [-q+q^{2p}+q^{1+p}(-1-q+2q^{p}-q^{2p})+q^{2(1+p)}(2+q-q^{p}-q^{2p})\cr
&~~~+ Q( 1+q^p-q^{2p}+q^{1+p}(1-2q^{p})+q^{2(1+p)}(-2-q+q^{p})+q^{3(1+p)}(1+q^{p}))], \cr
\widehat f^{c=1}_{\tableau{2 1}} =& {(-1)^p q^{-3p} (-1+q^{1+p}) Q^{1/2} \over(-1+q)^4 (1+q)^2 (1+q^2+q^4)} \cr
&\times [q (-1+q^p) (1+q^p+q^{1+p}(1-2q^{p})+q^{2(1+p)}(-2-2q)+q^{3(1+p)}(1+q)+q^{4(1+p)})\cr
& ~~~-Q (1+q) (-1 +q^{1+p}(-1 +3q^{p})+ q^{2(1+p)}(2+3 q - 3 q^{p})\cr
&~~~~~~~~~~~+q^{3(1+p)}(-2 -3q^{2})+q^{4(1+p)}+q^{5(1+p)})\cr
&~~~+ Q^2 (-1 +q^{1+p}(-1+2q^{p}) +q^{2(1+p)}(2+3q+q^{2}-q^{p} ) \cr
&~~~~~~~~~~~+ q^{3(1+p)}(-3-2q-2q^{2}) +q^{5+4p}+q^{6+5p})],\cr
\widehat f^{c=1}_{\tableau{1 1 1}} =&{(-1)^p q^{-1-3p} (1-q^{1+p}-q^{2+p}+q^{3+2p}) Q^{1/2}(-1+Q) \over(-1+q)^4 (1+q)^2 (1+q^2+q^4)} \cr
&\times [-q (1+q^{1+p}(1+q-q^{p})+q^{2(1+p)}(-2-2q-q^{2})+q^{3(1+p)}(1+q)+q^{5+4p}))\cr
&~~~+Q(1+q^{1+p}(1+q)+q^{2(1+p)}(-1-2q-2q^{2}-q^{3})+q^{3(1+p)}(q^{2}+q^{3})+q^{7+4p})].
}
$$
One can indeed check that, for any integer $p$, the above polynomials are 
of the form predicted in \cs\ (they are polynomials in $(q^{1/2}-q^{-1/2})^2$ 
with integer coefficients).

\subsec{$\IP^2$ attached to ${\IR \IP^2}$}

\ifig\geounorptwo{A D-brane in an outer leg of the orientifold of the two ${\IP^2}$'s connected by a ${\IP^1}$ .}
{\epsfxsize 5.5truein\epsfbox{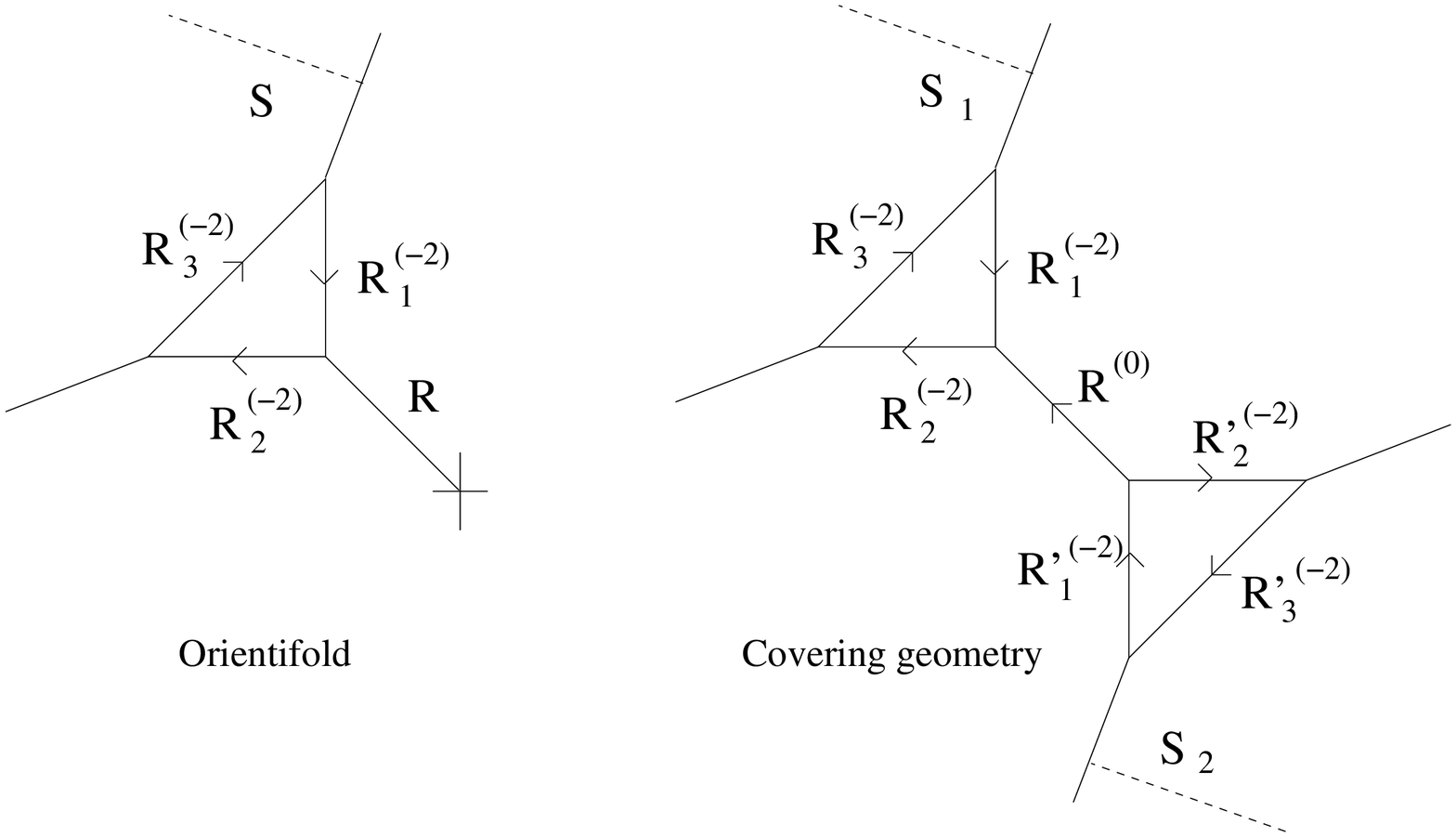}}

The next example we consider is the orientifold studied in \refs{\DFM,\BFM}, with a D-brane located in an outer leg. In this case, the covering space consists of two ${\IP^2}$'s connected by a ${\IP^1}$, with two 
D-branes in opposite legs  (the geometry is shown in \geounorptwo). Let us now define the following operator corresponding to the ${\IP^2}$ with an outer D-brane:
$$
\CO_{R S} = \sum_{R_i} q^{\sum_i \kappa_{R_i}} (-1)^{\sum_i \ell (R_i)} C_{S R_3 R_1^T} C_{\cdot R_2 R_3^T} C_{R_1 R_2^T R} e^{-t \sum_i \ell(R_i)}
$$
where $S$ is the representation attached to the D-brane. Using the topological vertex rules we can write,
for arbitrary framing $p$ ($Z_{\rm closed}$ is the amplitude without D-branes): 

$$
{\cal C}_{S_1 S_2}= {1 \over Z_{\rm closed}^{\rm cov}} \sum_{R}(-1)^{p(\ell (S_1) + \ell (S_2))} q^{{p \over 2} (\kappa_{S_1}+\kappa_{S_2})} \CO_{R S_1} \CO_{R^T S_2} (-Q)^{\ell(R)}
$$
and
\eqn\Ptwocovclosed{
Z_{\rm closed}^{\rm cov} = 1 + \sum_{R} \CO_{R \cdot} \CO_{R^T \cdot} (-Q)^{\ell(R)},
}
where in the last equation we have singled out the term where all the representations are trivial. As we do not have a closed expression for 
${\cal C}_{S_1 S_2}$ we have to evaluate $Z_{\rm or}$ order by order in $Q$ and $e^{-t}$.

Let us compute now $Z_S$ by using the topological vertex rules for orientifolds developed in \BFM. We find that
$$
Z_S = {1 \over Z_{\rm closed}} (-1)^{p \ell(S)} q^{p \kappa_S \over 2} \sum_{R=R^T} \CO_{R S} Q^{\ell(R)/2} (-1)^{{1\over 2} (\ell(R) \mp r(R))}
$$
where
\eqn\Ptwoclosed{
Z_{\rm closed} = 1+ \sum_{R=R^T} \CO_{R \cdot} Q^{\ell(R)/2} (-1)^{{1\over 2} (\ell(R) \mp r(R))}.
}
Using the results in the previous section, we can compute the functions ${\widehat f_S^c} (q, Q, e^{-t})$. 
We find the following results at low order, for arbitrary framing $p$ (again we present the results for $SO(N)$):
$$
\eqalign{
\widehat f^{c=1}_{\tableau{1}}=& (-1)^p \Bigl[ -Q^{1/2} e^{-t} + 4 Q^{1/2} e^{-2 t} -(35 + 8  \, z)Q^{1/2} e^{-3t} \cr
& \,\,\, \,\,\,\,\,\, + (400 + 344\, z 
+ 112 \, z^2 + 13 \, z^3)Q^{1/2} e^{-4t} -2  Q^{3/2} e^{-2 t} + (30 + 6  \, z)Q^{3/2} e^{-3t}\cr
& \,\,\,\,\,\,\,\,\,- (488 + 359\, z + 104 \, z^2 + 11 \, z^3)Q^{3/2} e^{-4t} - 3\, Q^{5/2} e^{-3t} \cr
& \,\,\,\,\,\,\,\,\,+ (132 + 59\, z + 8 \, z^2 )Q^{5/2} e^{-4t} + \cdots \Bigr],\cr
\widehat f^{c=2}_{\tableau{1}}=& -(-1)^p \Bigl[Q^2 e^{-3t} -(15 + 7 \, z + z^2) Q^2 e^{-4t} + 2\, Q^4 e^{-4t}+ \cdots\Bigr] ,\cr
\widehat f^{c=1}_{\tableau{2}}=& {q^{-p+1} (-1+q^{p}) (-q+q^{p}) \over (q-1)^2 (q+1)} Q^{1/2} e^{-t} 
- {3  q^{-p+1} (-1+q^{p})^2 \over (q-1)^2} Q^{1/2} e^{-2t} + \cdots, \cr
\widehat f^{c=2}_{\tableau{2}}=&{q^{-p+1} (q^p-1)^2 \over (q-1)^2} Q^2 e^{-3t} + \cdots,\cr
\widehat f^{c=1}_{\tableau{1 1}}=& {q^{-p+1} (q^p-1) (q^{1+p}-1) \over (q-1)^2 (q+1)} Q^{1/2} e^{-t}-{3 q^{-p} (q^{1+p}-1)^2 \over (q-1)^2} Q^{1/2} e^{-2t} 
+ \cdots, \cr
\widehat f^{c=2}_{\tableau{1 1}}=&{q^{-p} (q^{1+p}-1)^2 \over (q-1)^2} Q^2 e^{-3t} + \cdots ,\cr
}
$$
where $z\equiv (q^{1\over 2}- q^{-{1\over 2}})^2$. 

By comparing with \cs, it is easy to see that the results above have the expected polynomial form with integer coefficients for any $p$. In contrast to the 
example above of a D-brane in the orientifold of the conifold, in this example there are nonzero amplitudes with an even number of 
crosscaps. 

\subsec{$SO$/$Sp$ Hopf link invariant}

\ifig\geohopf{Two adjacent D-branes in the outer legs of the orientifold of 
the resolved conifold.}
{\epsfxsize 5.5truein\epsfbox{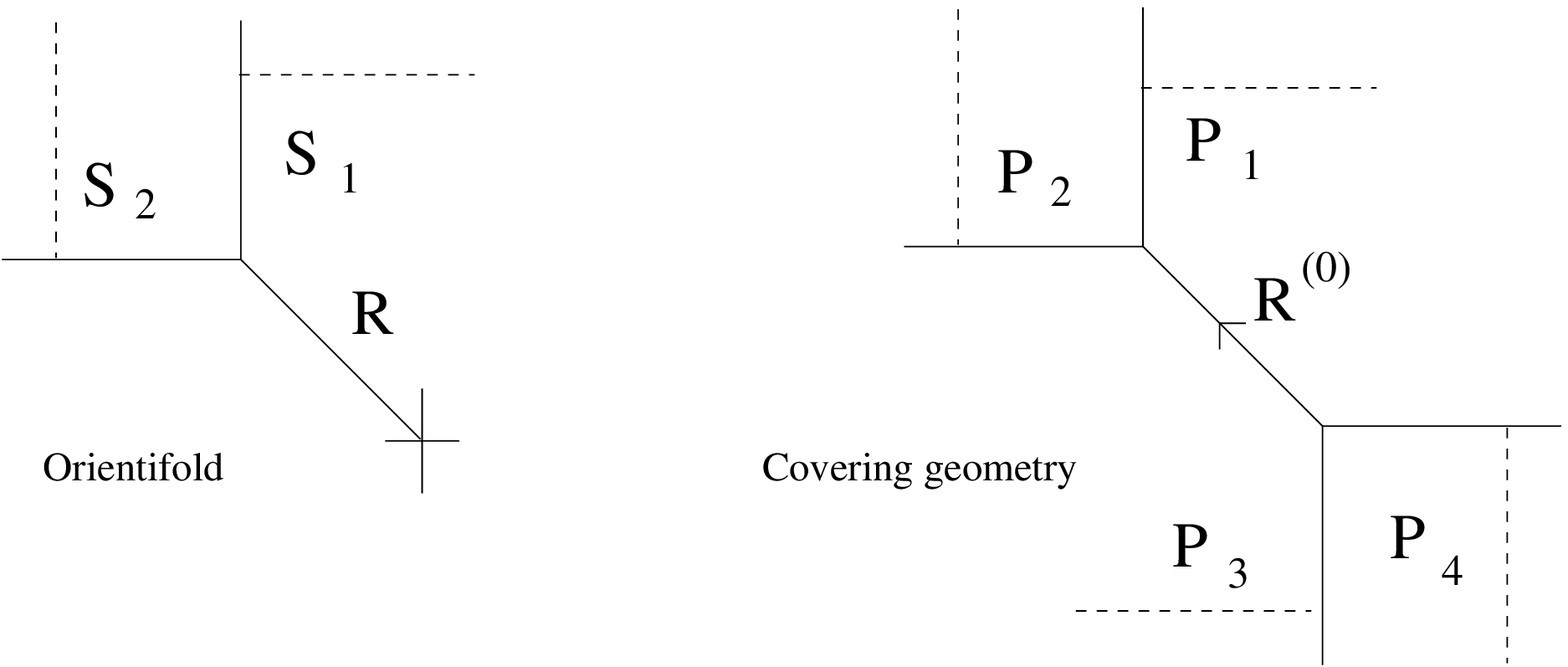}} 

 Our third and final example is the orientifold of the resolved conifold with two adjacent D-branes in the outer legs. The covering geometry now involves 
four sets of D-branes in the outer legs of the resolved conifold, oppositely identified by the involution. The geometry is shown in \geohopf.

Using the topological vertex, we find for the covering amplitude (for arbitrary framings $p_1$ and $p_2$):
\eqn\covampiii{\eqalign{
{\cal C}_{P_1 P_2 P_3 P_4} =&{1 \over Z_{\IP^1}} \sum_R (-1)^{p_1 (\ell(P_1)+\ell(P_3)) + p_2 (\ell(P_2)+\ell(P_4))} \cr
&\times q^{{p_1 \over 2} (\kappa_{P_1}+\kappa_{P_2})+{p_2 \over 2} (\kappa_{P_2}+\kappa_{P_4})} (-Q)^{\ell(R)} C_{R^T P_1 P_2} C_{R P_3 P_4}
}}
To obtain the oriented amplitude from \covampiii\ we have to identify the moduli $P_1$ ($P_2$) with $P_3$ ($P_4$) as explained in \zrorL. 
We can rewrite \covampiii\ by using the expression of the topological vertex in terms of Schur functions \topvertexschur:
\eqn\covampschur{\eqalign{
{\cal C}_{P_1 P_2 P_3 P_4} =&{1 \over Z_{\IP^1}} q^{{1\over2}(\sum_{i=1}^4 \kappa_{R_i})} s_{R_1^T} (q^{\rho}) s_{R_3^T} (q^{\rho}) \sum_{\eta_1, \eta_2} (-Q)^{\ell(\eta_1)} s_{R_2^T/\eta_1} (q^{\ell(R_1)+\rho}) s_{R_4^T/\eta_2} (q^{\ell(R_3)+\rho})\cr
& \times \sum_{R} s_{R^T/\eta_1} (-Q q^{\ell(R_1^T)+\rho}) s_{R/\eta_2} (q^{\ell(R_3^T)+\rho}).
}}
By using the identities \identitytwo, \identitythree\ and \identityfour\ we finally obtain that
\eqn\covampfinal{\eqalign{
{\cal C}_{P_1 P_2 P_3 P_4} =& q^{{1\over2}(\sum_{i=1}^4 \kappa_{R_i})} 
s_{R_1^T} (q^{\rho}) s_{R_3^T} (q^{\rho}) \prod_k (1-Q q^k)^{C_k (R_1^T, R_3^T)}\cr
&\sum_{\eta} (-Q)^{\ell(\eta)} s_{R_2^T/\eta^T} (q^{\ell(R_1)+\rho}, Q q^{-\ell(R_3^T)-\rho}) s_{R_4^T/\eta} (q^{\ell(R_3)+\rho}, Q q^{-\ell(R_1^T)-\rho}),
}}
where we defined the functions
\eqn\schurdouble{
s_{R/Q} (x,y) = \sum_{\eta} s_{R/\eta} (x) s_{\eta/Q} (y),
}
and the coefficients $C_k (R_1^T, R_3^T)$ are defined in \expcoeff\ or \coeffs. Notice that \covampfinal\ is a polynomial in $Q$.

Now that we have our final expression for the covering amplitude, let us look at the full amplitude. The vertex rules for orientifolds developed in our previous paper \BFM\ tell us that, for the amplitude where there are two D-branes in the outer legs, one has (for arbitrary framings $p_1$ and $p_2$)
\eqn\amptwofinal{\eqalign{
Z_{S_1 S_2} &={1 \over Z_{X/I}} \sum_{R=R^T}(-1)^{p_1 \ell(S_1)+p_2 \ell(S_2)} q^{{1 \over 2}(p_1 \kappa_{S_1}+p_2 \kappa_{S_2})} C_{S_1 S_2 R} Q^{\ell (R)/2}(-1)^{{1\over 2}(\ell(R)\mp r(R))} \cr
 &=(-1)^{p_1 \ell(S_1)+p_2 \ell(S_2)} q^{{1 \over 2}(p_1 \kappa_{S_1}+p_2 \kappa_{S_2})}q^{{1 \over 2} \kappa_{S_2}} Q^{{1 \over 2} (\ell(S_1)+\ell(S_2))} \CW_{S_1 S_2^t}^{SO(N)/Sp(N)},
}}
where we used again \magic. This time, we see that the {\it total} amplitude of the two D-brane configuration in the orientifold of the conifold is given by the $SO/Sp$ Chern-Simons invariants of the Hopf link.

By substracting the oriented piece from the unoriented amplitude, and using the results of the previous section, we can compute the $N^{c}_{(S_1, S_2), g, \beta}$ integer invariants through the $\widehat f_{S_1 S_2}^c$ functions. As noted in \tildefrlinks\ we now expect a slightly different structure for the $\widehat f_{S_1 S_2}^c$ functions than the one given by \cs, since $L=2$. Namely, we expect
\eqn\fhattwo{
\widehat f_{S_1 S_2}^c = \sum_{g, \beta} N^{c}_{(S_1, S_2), g, \beta} (q^{1\over 2} - q^{-{1\over 2}})^{2g+1} Q^{\beta}.
}
with the $\widehat f_{S_1 S_2}^c$ functions defined as in \ccont.

We obtain the following results for $SO(N)$:
\eqn\resulthopf{\eqalign{
{\widehat f}_{\tableau{1} \tableau{1}}^{c=1} &= (-1)^{p_1+p_2} Q^{1/2} (1-Q) (q^{1/2}-q^{-1/2}),\cr
{\widehat f}_{\tableau{2} \tableau{1}}^{c=1} &={(-1)^{2p_1+p_2} Q^{1/2} (1-Q) q^{-p_1-1/2} \over q-1} (-q+2q^{1+p_1}-q^{1+2p_1} +Q(q^2 +q^{2p_1} -2q^{1+p_1})),\cr 
{\widehat f}_{\tableau{1 1} \tableau{1}}^{c=1}&={(-1)^{2p_1+p_2} Q^{1/2} (1-Q) q^{-p_1-1/2} \over q-1}(-(-1+q^{1+p_1})^2 + Q q (-1 + q^{p_1})^2).
}}
It is straigthforward to show that for any fixed framings $p_1$ and $p_2$ the $\widehat f$ functions \resulthopf\ have the structure predicted by \fhattwo\ with integer invariants $N^{c}_{(S_1, S_2), g, \beta}$. Up to the order $\ell(S_1)+\ell(S_2) =3$ the contributions with two crosscaps vanish.

\subsec{Localization computations}

In the previous subsections we found many open BPS invariants using the topological vertex prescription of \BFM\ and the structure predictions of section \structure. As far as we are aware these invariants have never been computed before. Therefore it would be nice to have an independent check of our results which does not rely on large $N$ duality.

In \refs{\DFM, \BFM} localization techniques were defined to compute closed unoriented Gromov-Witten invariants of Calabi-Yau orientifolds. In this section we will extend these techniques to the case of {\it open} unoriented Gromov-Witten invariants, therefore providing an alternative and independent way to compute the invariants of the previous subsections.

In order to compare our results with localization computations we have to extract open Gromov-Witten invariants from the $f$ polynomials. First let us recall the definition of the $f$ functions \conj:
\eqn\deffunct{
{\cal F}^c (V_1, \dots, V_L) = - \sum_{d=1}^\infty \sum_{R_1, \cdots, R_L}
{1\over d} f_{(R_1, \cdots, R_L)}^c (q^d, {\rm e}^{-d t})
\prod_{\alpha=1}^L {\rm Tr}_{R_\alpha}V_{\alpha}^d
}
where we added the superscript $c$ for the number of crosscaps. As usual, we can also work in the $\vec{k}$ basis. In this basis the free energy reads (see \LMii):
\eqn\freekbasis{
{\cal F}^c (V_1, \dots, V_L) = - \sum_{\{ \vec{k}^{\alpha} \}} W_{(\vec{k}^{(1)}, \dots, \vec{k}^{(L)})}^{({\rm conn}),c} \prod_{\alpha}^L {1 \over z_{\vec{k}^{(\alpha)}}} \Upsilon_{\vec{k}^{(\alpha)}} (V_{\alpha}),
}
where we defined the connected vevs $W_{(\vec{k}^{(1)}, \dots, \vec{k}^{(L)})}^{({\rm conn}),c}$, 
and $z_{\vec{k}} = \prod_m k_m ! m^{k_m}$. Since $q=e^{i g_s}$, we can expand the r.h.s of \freekbasis\ in $g_s$. We find a series with the structure \LMii:
\eqn\fgwfunct{\eqalign{
{\cal F}^c (V_1, \dots, V_L) &= \sum_{g=0}^{\infty} i^{\sum_{\alpha=1}^L |{\vec{k}^{(\alpha)}}|+c} F_{g,(\vec{k}^{(1)}, \dots, \vec{k}^{(L)})}^c g_s^{2g-2+c+\sum_{\alpha=1}^L |{\vec{k}^{(\alpha)}}|} \Upsilon_{\vec{k}^{(\alpha)}} (V_{\alpha}) \cr
&=- {\left(  \prod_{\alpha}^L {1 \over z_{\vec{k}^{(\alpha)}}} \right)} W_{(\vec{k}^{(1)}, \dots, \vec{k}^{(L)})}^{({\rm conn}),c} \Upsilon_{\vec{k}^{(\alpha)}} (V_{\alpha}),
}}
where $F^c_{g,(\vec{k}^{(1)}, \dots, \vec{k}^{(L)})}$ is the generating functional for open Gromov-Witten invariants at genus $g$, with $c$ crosscaps and fixed boundary conditions given by $(\vec{k}^{(1)}, \dots, \vec{k}^{(L)})$. The factor of $i^{\sum_{\alpha=1}^L |{\vec{k}^{(\alpha)}}|+c}$ is necessary to compare Chern-Simons (or topological vertex) results with localization computations \MV. Thus, we see that to extract open Gromov-Witten invariants we have to compute the connected vevs 
$W_{(\vec{k}^{(1)}, \dots, \vec{k}^{(L)})}^{({\rm conn} ),c}$ from the $f$ functions. Such a relation has been deduced in \LMii:
\eqn\relvevf{
W_{(\vec{k}^{(1)}, \dots, \vec{k}^{(L)})}^{({\rm conn}),c} = \sum_{d| \vec{k}^{(\alpha)},~d~{\rm odd}} d^{\sum_{\alpha} | \vec{k}^{(\alpha)}|-1} \sum_{\{ R_{\alpha}\}} \prod_{\alpha=1}^{L} \chi_{R_{\alpha}} ( C(\vec{k}^{(\alpha)}_{1/d})) f_{(R_1, \dots, R_L)}^c (q^d, {\rm e}^{-dt}),
}
where $C(\vec{k})$ is the conjugacy class associated to a vector $\vec{k}$, which has $k_j$ cycles of length $j$, and $\chi_{R}$ is the character of the symmetric group $S_{\ell}$. In \relvevf\ the vector $\vec{k}_{1/d}$ is defined as follows. Fix a vector $\vec{k}$, and consider all the positive integers $d$ that satisfy the following condition: $d|j$ for every $j$ with $k_j \neq 0$. When this happens, we will say that ``$d$ divides $\vec{k}$", and we will denote this as $d|\vec{k}$. We can then define the vector $\vec{k}_{1/d}$ whose components are $(\vec{k}_{1/d})_i = k_{di}$. In \relvevf\ the integer $d$ has to divide all the vectors $\vec{k}^{(\alpha)}$, $\alpha=1,\dots,L$. Note that in \relvevf\ the sum is only over $d$ {\it odd}: this is because in the unoriented case only odd multicovers contribute.

Using \fgwfunct\ and \relvevf\ one can find expressions for the generating functionals of open Gromov-Witten invariants in terms of $f$ functions. Let us define the all genera generating functionals for open Gromov-Witten invariants with $c$ crosscaps and fixed boundary conditions given by $(\vec{k}^{(1)}, \dots, \vec{k}^{(L)})$:
\eqn\allgene{
F_{(\vec{k}^{(1)}, \dots, \vec{k}^{(L)})}^c = \sum_{g=0}^{\infty} F_{g,(\vec{k}^{(1)}, \dots, \vec{k}^{(L)})}^c g_s^{2g-2+c+\sum_{\alpha=1}^L |{\vec{k}^{(\alpha)}}|}.
}
For configurations with one representation ($L=1$), one finds
\eqn\genfunct{\eqalign{
&F^c_{(1,0,\dots)} =  i^{1-c} f^c_{\tableau{1}}, ~~~~~~~~F^c_{(2,0,\dots)} = {i^{-c} \over 2} (f^c_{\tableau{2}}+f^c_{\tableau{1 1}}),~~~~~~~~F^c_{(0,1,0,\dots)}= {i^{1-c} \over 2} (f^c_{\tableau{2}}-f^c_{\tableau{1 1}})\cr
&~~~~~F^c_{(3,0,\dots)}=-{i^{1-c} \over 6} (f^c_{\tableau{3}}+2 f^c_{\tableau{2 1}}+f^c_{\tableau{1 1 1}}),~~~~~~~~~~~~F^c_{(1,1,0,\dots)}={i^{-c} \over 2} (f^c_{\tableau{3}}-f^c_{\tableau{1 1 1}})\cr
&~~~~~~~~~~~~~F^c_{(0,0,1,0,\dots)}={i^{1-c} \over3}(f^c_{\tableau{3}}- f^c_{\tableau{2 1}}+f^c_{\tableau{1 1 1}}+ f^c_{\tableau{1}} (q^3, {\rm e}^{-3t})),
}}
For configurations with two representations ($L=2$), one finds
\eqn\genfuncttwo{\eqalign{
&F^c_{((1,0,\dots),(1,0,\dots))} = i^{-c}f^c_{\tableau{1} \tableau{1}},~~~~~~~~F^c_{((2,0,\dots),(1,0,\dots))} = -{i^{1-c} \over2}(f^c_{\tableau{2} \tableau{1}}+f^c_{\tableau{1 1} \tableau{1}}),\cr
&~~~~~~~~~~~~~~~F^c_{((0,1,0,\dots),(1,0,\dots))} = {i^{-c}\over2}(f^c_{\tableau{2} \tableau{1}}-f^c_{\tableau{1 1} \tableau{1}}).
}}

Using the above formulae, we can compute the $F^c_{(\vec{k}^{(1)}, \dots, \vec{k}^{(L)})}$ generating functionals and put them in the form 
of \allgene\ by expanding in $g_s$. This will extract the open Gromov-Witten invariants from our previous results. 

\eject

\vskip 14pt

{\bf I. The $SO/Sp$ framed unknot}

\vskip 6pt

\noindent The topological vertex gives the following results:
$$
\eqalign{
F_{(1,0,0,...)}^{c=1} =& (-1)^p Q^{1/2},\cr
F_{(2,0,0,...)}^{c=1}=& {{1\over 2}\left[(1+p)^2 Q^{1/2} (1-Q)\right]} g_s  - {{1 \over 48}\left[  (1+p)^2 (1+4p+2p^2) Q^{1/2} (1-Q)\right]} g_s^3 +\dots,\cr
F_{(0,1,0,...)}^{c=1}=& {\left[(1+p) Q^{1/2} (1-Q)\right]}-{{1 \over 24} \left[(3+11p+12p^2+4p^3) Q^{1/2} (1-Q)\right]} g_s^2 +\dots,\cr
F_{(3,0,0,\dots)}^{c=1}=& { {1\over6}\left[(-1)^p Q^{1/2} (1+p)^3  (1+3p-6Q(1+p)+Q^2 (5+3p))\right]} g_s^2 + \dots,\cr
F_{(1,1,0,\dots)}^{c=1} =& {\left[ (-1)^p Q^{1/2} (1+p)^2 (1+2p - 4Q(1+p)+Q^2(3+2p)) \right]} g_s + \dots,\cr
F_{(0,0,1,\dots)}^{c=1} = &{ {1\over6}\left[ (-1)^p Q^{1/2}(3(1+p)(2+3p+Q^2(4+3p))-2Q(8+18p+9p^2))\right]}+\dots~.
}
$$

In order to compare with the localization computation, we introduce first some notation. We will consider the following real torus 
action on the resolved conifold $X$:
$$
e^{i\phi}\cdot(X_1,X_2,X_3,X_4)=(e^{i\lambda_1\phi}X_1,e^{i\lambda_2\phi}X_2,e^{i\lambda_3\phi}X_3,e^{i\lambda_4\phi}X_4).
$$
The weights of the torus action on the local coordinates $z=X_1/X_2,u=X_2X_3,v=X_2X_4$ are given by $\lambda_z=\lambda_1-\lambda_2,
\lambda_u=\lambda_2+\lambda_3,\lambda_v=\lambda_2+\lambda_4$ respectively. Note that from the compatibility of the torus action with the 
antiholomorphic involution it follows that $\lambda_u+\lambda_v+\lambda_z=0$. Now we can present the localization results:
$$
\eqalign{
F_{(1,0,0,...)}^{c=1} =& Q^{1\over 2},\cr
F_{(2,0,0,...)}^{c=1}=& {1\over 2}\left[\left({a\over a-1}\right)^2 Q^{1\over 2}(1-Q)\right] g_s-{1 \over 48}\left[{a^2(a^2+2a-1)\over (a-1)^4} 
Q^{1\over 2}(1-Q)\right] g_s^3 +\dots,\cr
F_{(0,1,0,...)}^{c=1}=& \left[{a\over a-1} Q^{1\over 2}(1-Q)\right] -{1 \over 24}\left[{a(a+1)(3a-1)\over (a-1)^3} 
Q^{1\over 2}(1-Q)\right] g_s^2 +\ldots,\cr
F_{(3,0,0,\dots)}^{c=1}=& -{ {1\over6}\left[Q^{1/2} \left(a\over a-1\right)^3  \left( {a+2\over a-1}-{6a \over a-1}Q+\ldots\right)\right]}
g_s^2 + \dots,\cr
F_{(1,1,0,\dots)}^{c=1}=& -{\left[Q^{1/2} \left(a\over a-1\right)^2 \left( {a+1\over a-1}-{4 a \over a-1}Q+\ldots\right)\right]}
g_s + \dots,\cr
F_{(0,0,1,\dots)}^{c=1}=& -{ {1\over6}\left[Q^{1/2}\left( {3 a(2a+1)\over (a-1)^2}-{2(8a^2+2a-1) \over (a-1)^2}Q +\ldots\right)
\right]} + \dots~.
}
$$
where $a=-\lambda_v/\lambda_z$. After making the substitution $a=1+{1\over p}$, we find that the above results coincide with the 
expressions obtained from the vertex computation up to factors of $\pm (-1)^p$. The sign difference is due to 
different choice of conventions between the vertex and the localization computations. 
As an example, we present below the graphs contributing to the unoriented open 
Gromov-Witten invariant for genus $1$ maps with degree $3$ $\IR\IP^2$ and winding vector $(0,1,0,...)$, as well as their contributions.

\ifig\unveriv{One crosscap, genus $1$ and three crosscaps, genus $0$ at degree $3$ $\IR\IP^2$, and winding vector $(0,1,0,...)$.}
{\epsfxsize5.2in\epsfbox{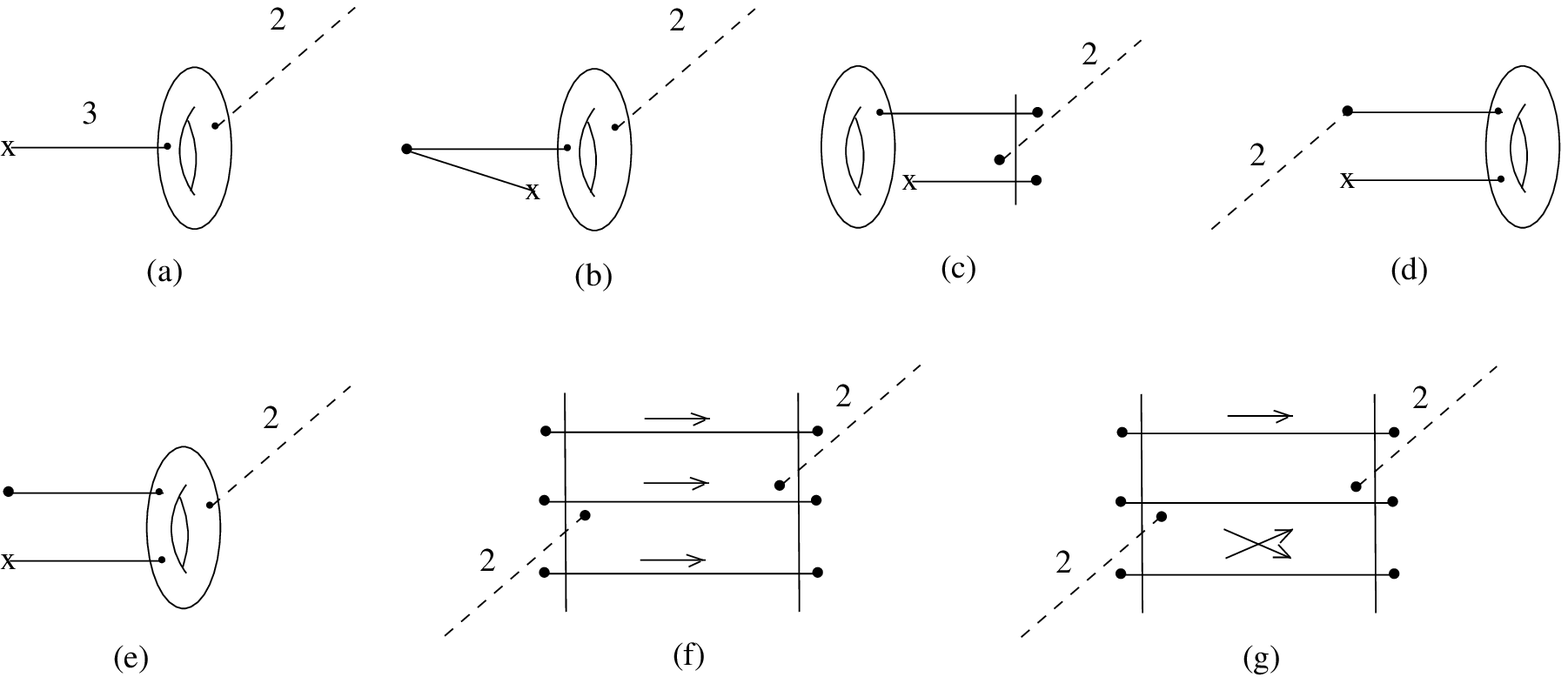}}

\noindent The contributions of the above graphs are computed according to the rules explained in \DFM. We obtain:
$$
\eqalign{
& C_{(0,1,0,...)}^{(1,3),(a)}={(\lu-2\lv)(2\lu-\lv)\lv(\lu+2\lv)(\lambda_u^3+6\lambda_u^2\lv+\lu\lambda_v^2+2\lambda_v^3)
\over 48\lambda_u^3\lambda_z^4},\cr
& C_{(0,1,0,...)}^{(1,3),(b)}=-{(\lu-2\lv)\lambda_v^2(\lu+2\lv)\over 48\lambda_u^2\lambda_z^2},\quad C_{(0,1,0,...)}^{(1,3),(c)}
={\lambda_v^2(\lu+2\lv)
\over 24\lu\lambda_z^2},\cr
& C_{(0,1,0,...)}^{(1,3),(d)}=-{\lambda_v^2(2\lu+\lv)(\lu+2\lv)\over 24\lambda_z^4},\quad C_{(0,1,0,...)}^{(1,3),(e)}={\lambda_v^3
(\lu+2\lv)(\lambda_u^3-2\lu\lambda_v^2-2\lambda_v^3)\over 12\lambda_u^3 \lambda_z^4},\cr
& C_{(0,1,0,...)}^{(1,3),(f)}=-{(\lu+2\lv)\lambda_v^3(\lu-2\lv)\over 6\lu\lambda_z^4},\quad  
C_{(0,1,0,...)}^{(1,3),(g)}={(\lu+2\lv)\lambda_v^3(\lu-2\lv)\over 2\lu\lambda_z^4}.
}
$$

\vfill\eject


{\bf II. $\IP^2$ attached to $\IR\IP^2$}

\vskip 6pt

\noindent The topological vertex gives the following results:
$$
\eqalign{
F_{(1,0,...)}^{c=1}=& (-1)^p Q^{1/2} e^{-t} [-1 -2(-2+Q)e^{-t} +(-35+30 Q-3Q^2) e^{-2t}\cr
&+ 4(100-122Q+33Q^2)e^{-3t}+\dots]+\dots, \cr
F_{(1,0,...)}^{c=2}=& -(-1)^p Q^2 e^{-3t} {\left[ 1+(-15+2Q^2) e^{-t}+\dots \right]} g_s +\dots, \cr
F_{(2,0,...)}^{c=1}=& {1 \over 2} Q^{1/2} e^{-t} {\left[-p^2 + (3 + 6p + 6p^2) e^{-t} +\dots \right]} g_s+\dots, \cr
F_{(2,0,...)}^{c=2}=& -{1 \over 2} Q^2 e^{-3t} {\left[ 1+2p+2p^2+\dots \right]} g_s^2 + \dots,\cr
F_{(0,1,0,...)}^{c=1}=&Q^{1/2} e^{-t} {\left[ -p + (3+6p)e^{-t}+\dots \right]} + \dots, \cr
F_{(0,1,0,...)}^{c=2}=& -Q^2 e^{-3t} {\left[ 1+2p +\dots \right]} g_s  + \dots~.
}
$$

\noindent For the localization computations, we will use the same notation as in \BFM. We present below some of the localization 
computations we performed. First, we obtain
$$
\eqalign{
F_{(2,0,...)}^{c=1}=&{1\over 2} Q^{1/2} e^{-t} {\left[-\left({\lv-\lu\over\lv-2\lu}\right)^2 + \left({3 (2 \lambda_u^2-2\lu\lv + \lambda_v^2)\over 
(2\lu-\lv)^2}\right) e^{-t} +\dots \right]} g_s 
+\dots~.}
$$
We present the graphs contributing at degree $2$ hyperplane class in the expression above, as well as their contributions.
\ifig\unveriv{One crosscap graphs at degree $1$ $\IR\IP^2$, degree $2$ hyperplane and winding vector $(2,0,...)$.}
{\epsfxsize5.5in\epsfbox{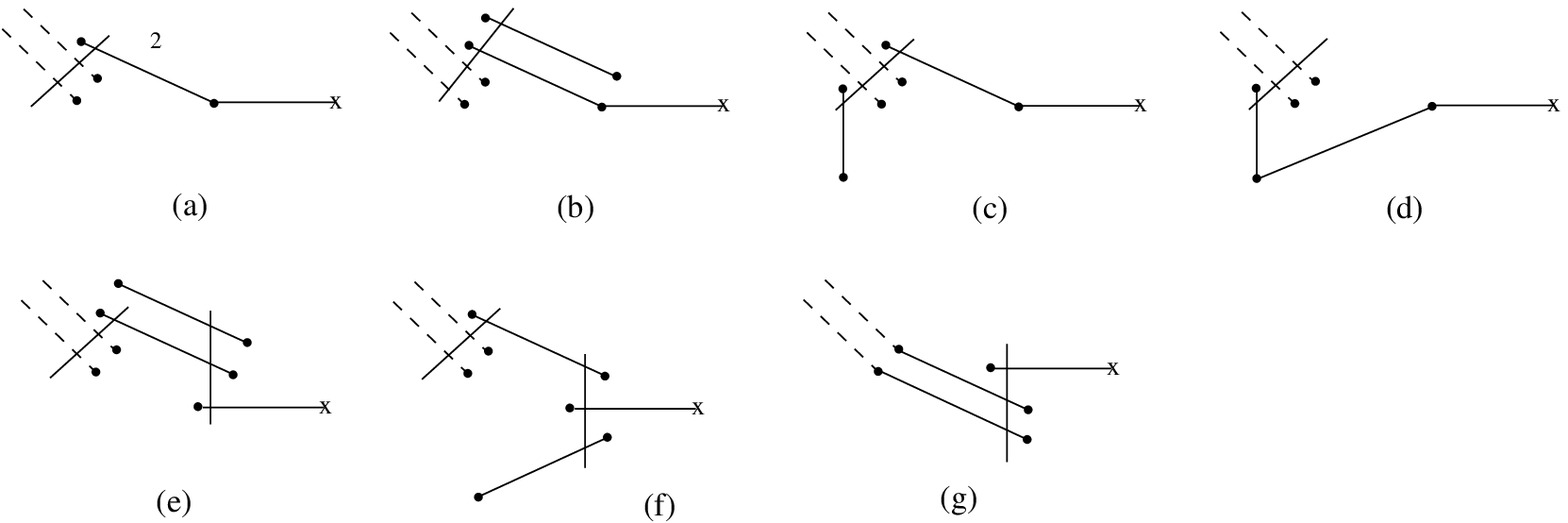}}
$$
\eqalign{
& C_{(-1,1,2)}^{(2,0,0,...),(a)}={\lv(\lu-\lv)^2(3\lu-2\lv)\over \lambda_u^2(\lv-2\lu)^2},\quad 
C_{(-1,1,2)}^{(2,0,0,...),(b)}={(\lu-\lv)^4\over \lambda_u^2(\lv-2\lu)^2},\cr
&  C_{(-1,1,2)}^{(2,0,0,...),(c)}={2\lambda_u^2-2\lu\lv+\lambda_v^2\over 2(\lv-2\lu)^2},\quad
 C_{(-1,1,2)}^{(2,0,0,...),(d)}={\lambda_u^2\over 2(\lv-2\lu)^2},\cr
& C_{(-1,1,2)}^{(2,0,0,...),(e)}={\lambda_v^2(\lu-\lv)^2\over 2\lambda_u^2(\lv-2\lu)^2},\quad
C_{(-1,1,2)}^{(2,0,0,...),(f)}={(\lu-\lv)^2\over 2(\lv-2\lu)^2}, \quad C_{(-1,1,2)}^{(2,0,0,...),(g)}={\lambda_v^2\over 2\lambda_u^2}.
}
$$
\noindent In the expressions above, the subscript of the contributions is $(\chi,d_1,d_2)$ where $\chi$ is the unoriented genus of the 
closed component of the map and $d_1$ and $d_2$ are the $\IR\IP^2$ and hyperplane degrees respectively. Then, for the same winding vector, 
at $2$ crosscaps we obtain 
$$
F_{(2,0,...)}^{c=2}=-{1\over 2}Q^2e^{-3t}\left[{2\lambda_u^2-2\lu\lv+\lambda_v^2\over (\lv-2\lu)^2}+\dots\right]+\dots~.
$$

\noindent The two crosscaps configurations were discussed at length in \BFM. The graphs come in sets and there is a single set such that 
the sum of the contributions of the corresponding graphs does not vanish. That set and the graphs contributions are presented below.

\ifig\unveriv{Two crosscaps graphs at degree $4$ $\IR\IP^2$, degree $3$ hyperplane and winding vector $(2,0,...)$.}
{\epsfxsize5.2in\epsfbox{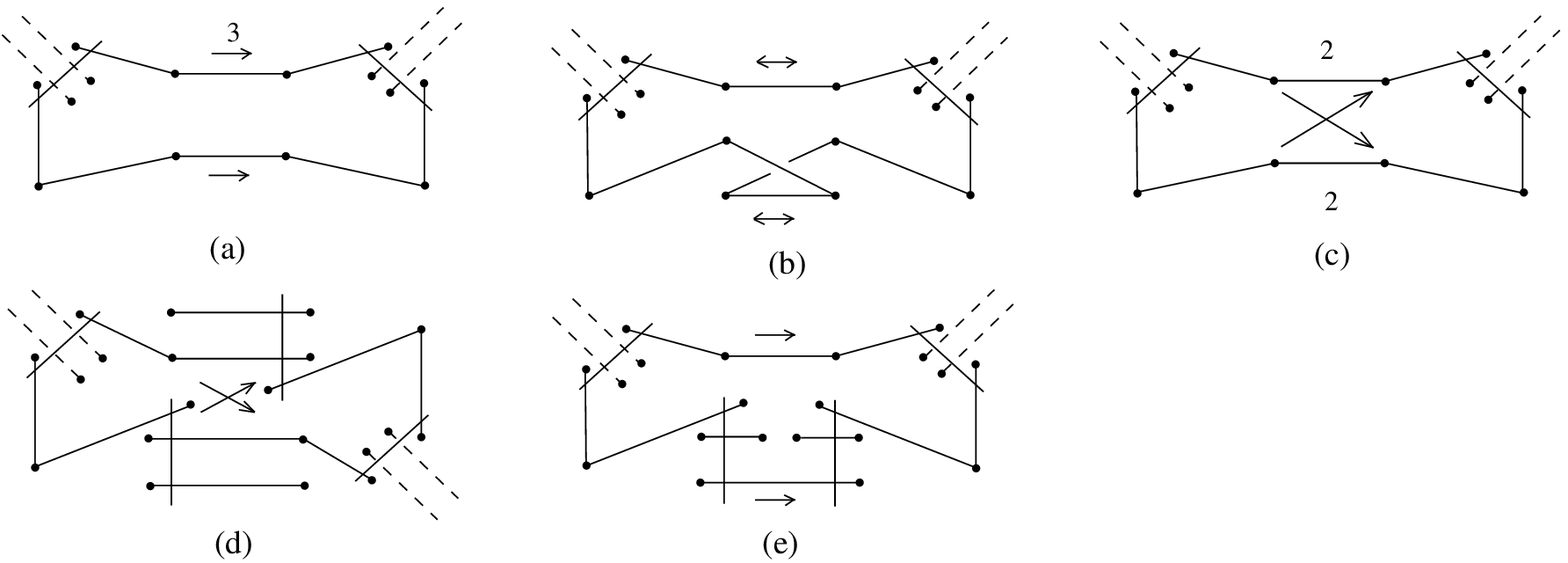}}
$$
\eqalign{
&C_{(0,4,3)}^{(2,0,0,...),(a)}=-{(\lambda_u^2-\lu\lv+\lambda_v^2)(2\lambda_u^2-2\lu\lv+\lambda_v^2)\over 2\lambda_z^2(\lv-2\lu)^2},\cr
&C_{(0,4,3)}^{(2,0,0,...),(b)}={1\over 2}C_{(0,4,3)}^{(2,0,0,...),(c)}=
-{\lu\lv(2\lambda_u^2-2\lu\lv+\lambda_v^2)\over 2\lambda_z^2(\lv-2\lu)^2},\cr
&C_{(0,4,3)}^{(2,0,0,...),(d)}=-C_{(0,4,3)}^{(2,0,0,...),(e)}=
{(\lambda_u^2+\lambda_v^2)(2\lambda_u^2-2\lu\lv+\lambda_v^2)\over 2\lambda_z^2(\lv-2\lu)^2}.
}
$$
We also obtain
$$
\eqalign{
&F_{(0,1,0,...)}^{c=1}=Q^{1/2}e^{-t}\left[{\lv-\lu\over \lv-2\lu}-3\left({\lv\over \lv-2\lu}\right)e^{-t}+\dots\right]+\dots,\cr
&F_{(0,1,0,...)}^{c=2}=Q^2e^{-3t}\left[\left({\lv\over \lv-2\lu}\right)+\dots\right]+\dots~.
}
$$
We note that for this geometry we obtain agreement with the vertex computation if we set $p=-{\lv-\lu\over\lv-2\lu}$.

\vskip 14pt

{\bf III. $SO/Sp$ Hopf link invariant}

\vskip 6pt

\noindent The results obtained from the topological vertex are:
$$
\eqalign{
F_{((1,0,...),(1,0,\dots))}^{c=1}=& (-1)^{p_1+p_2} Q^{1/2} (1-Q) g_s - {1 \over 24} (-1)^{p_1+p_2}Q^{1/2} (1-Q) g_s^3 + \dots, \cr
F_{((2,0,...),(1,0,\dots))}^{c=1}=& {1\over 2}(-1)^{p_2} Q^{1/2} {\left[1+2p_1+2p_1^2-2Q(1+2p_1^2)+Q^2(1-2p_1+2p_1^2) \right]} g_s^2 + \dots, \cr
F_{((0,1,0,...),(1,0,\dots))}^{c=1}=& (-1)^{p_2} Q^{1/2} {\left[1+2p_1 -4Q p_1-Q^2(1-2p_1) \right]} g_s + \dots~.
}
$$
\noindent The localization results are:
$$
\eqalign{
F_{((1,0,...),(1,0,\dots))}^{c=1}=& Q^{1/2} (1-Q) g_s - {1 \over 24}Q^{1/2} (1-Q) g_s^3 + \dots, \cr
F_{((2,0,...),(1,0,\dots))}^{c=1}=& -{1\over 2}Q^{1/2} \bigg[{\lambda_u^2+2\lu\lv+2\lambda_v^2\over \lambda_u^2}-2Q \left(\lu^2+2\lv^2 \over \lu^2 \right) \cr
&~~~~~~~~~~+ Q^2 \left(\lambda_u^2-2\lu\lv+2\lambda_v^2\over \lambda_u^2 \right) \bigg] g_s^2 + \dots, \cr
F_{((0,1,0,...),(1,0,\dots))}^{c=1}=& Q^{1/2} {\left[{\lu+2\lv\over\lu} -4Q \left(\lv \over \lu \right) -Q^2 \left(\lu-2 \lv \over \lu \right) \right]} g_s+\dots~.
}
$$
\noindent To obtain agreement with the vertex result for this D-brane configuration, we need to set $p_1={\lv\over\lu}$. These computations 
offer strong evidence of the equivalence between the vertex computation and the localization on the moduli space of stable open unoriented 
maps.

\newsec{Application: the BPS structure of the coloured Kauffman 
polynomial}

\seclab\kauffman

One of the most interesting applications of the above results is the 
determination of the BPS structure of the coloured Kauffman polynomial. 
In contrast to the results obtained for orientifolds of toric geometries above, 
we won't be able to give a full determination of all quantities involved for 
arbitrary knots, but we can still formulate some interesting structural 
properties of the knot polynomials similar to those explained in \refs{\OV,\LMi,\LMV,\LMii}. 
We will first recall the results for the coloured HOMFLY polynomial, and then 
we will state and illustrate the results for the coloured Kauffman polynomial. 

\subsec{Chern-Simons invariants and knot polynomials}

Let us consider Chern-Simons theory on ${\bf S}^3$ with gauge group $G$. 
The natural operators in this theory are the holonomies of the gauge 
connection around a knot ${\cal K}$, 
\eqn\wloop{
W_R^{\cal K}(A)={\rm P} \, \exp \oint_{\cal K} A.}
If we now consider a link ${\cal L}$ with components ${\cal K}_{\alpha}$, $\alpha=1,
\cdots, L$, the correlation function
\eqn\vevwilson{
W_{R_1 \cdots R_L}^G({\cal L})=\langle W^{{\cal K}_1}_{R_1}\cdots
W^{{\cal K}_L}_{R_L}\rangle}
defines a topological
invariant of the link ${\cal L}$. In this equation 
the bracket denotes a normalized vacuum expectation value, and we have 
indicated the gauge group $G$ as a superscript. It is well-known \W\ that 
Chern-Simons produces in fact invariants of {\it framed} links, but in the following 
we will consider knots 
in the so-called standard framing (see \refs{\guada,\MM} for a review of these topics). 
The correlation functions \vevwilson\ turn out to be rational functions 
of the variables $q^{\pm 1/2}$, $\lambda^{\pm 1/2}$. The variable $q$ is 
defined as $q=e^{i g_s}$, where $g_s$ is the effective Chern-Simons coupling 
constant
\eqn\stringcou{
g_s= { 2\pi \over k + y},}
$k$ is the coupling constant of Chern-Simons theory, and 
$y$ is the dual Coxeter of the gauge group (therefore it is $N$ for $U(N)$, $N-2$ for $SO(N)$, and
$N+1$ for $Sp(N)$). The variable $\lambda$ is defined by
\eqn\lam{
\lambda=q^{N+a},}
where
\eqn\avalues{
a=\cases{0\,\,\,\,\,\,\,\,\,\,\,\,\,\,\,\,\,\,\,\,\,\,\,\,\,\,\, {\rm for
\,\,\,\,} U(N), \cr
-1\,\,\,\,\,\,\,\,\,\,\,\,\,\,\,\,\,\,\,\,\,\, {\rm for
\,\,\,\,} SO(N),\cr
1\,\,\,\,\,\,\,\,\,\,\,\,\,\,\,\,\,\,\,\,\,\,\,\,\,\,\, {\rm for
\,\,\,\,} Sp(N).}}
The vacuum expectation values of Wilson loops are related to link invariants 
obtained from quantum groups \W:

1) If $G=U(N)$ and $R_1=\cdots =R_L=\tableau{1}$, then
\eqn\homflyrel{
W^{U(N)}_{\tableau{1} \cdots \tableau{1}}({\cal L})=
\lambda^{{\rm lk}({\cal L})}
\biggl( {\lambda^{1\over 2} -\lambda^{-{1\over 2}}
\over q^{1\over 2} -q ^{-{1\over 2}}} \biggr) P_{\cal L}(q, \lambda)}
where $P_{\cal L}(q, \lambda)$ is the HOMFLY polynomial of ${\cal L}$ \HOMFLY, 
and ${\rm lk}({\cal L})$ is its linking number.

2) If $G=SO(N)$ and $R_1=\cdots =R_L=\tableau{1}$, then
\eqn\kaufmannrel{
W^{SO(N)}_{\tableau{1} \cdots \tableau{1}}({\cal L})=
\lambda^{{\rm lk}({\cal L})}
\biggl(1+ {\lambda^{1\over 2} -\lambda^{-{1\over 2}}
\over q^{1\over 2} -q ^{-{1\over 2}}} \biggr) F_{\cal L}(q, \lambda)}
where $F_{\cal L}(q, \lambda)$ is the Kauffman polynomial of ${\cal L}$ \K.

We will call $W_{R_1 \cdots R_L}^{U(N)}({\cal L})$ and $W_{R_1 \cdots R_L}^{SO(N)}({\cal L})$ 
the coloured HOMFLY and Kauffman polynomials of ${\cal L}$, respectively. Note that there is 
a slight abuse of 
language here, since these Chern-Simons correlation functions are not polynomials, but rather rational 
functions.

\subsec{BPS structure: statement and examples}
  
In \OV, Ooguri and Vafa extended the duality of \GV\ between Chern-Simons theory on ${\bf S}^3$ and 
topological strings on the resolved conifold by incorporating the correlation functions \vevwilson. 
We will consider the case of knots, although everything we will say has a straigthforward 
generalization to links. The results of \OV\ are the following: first, to any knot ${\cal K} \in {\bf S}^3$ one can associate 
a Lagrangian submanifold $S_{\cal K}$ in the resolved conifold. Moreover, 
the generating functional of knot invariants 
\eqn\genun{
Z_{U(N)}(V)=\sum_R  W_R^{U(N)}({\cal K}) {\rm Tr}_R \, V
}
where $V$ is a $U(M)$ matrix, is the partition function for open topological strings propagating on the resolved 
conifold and with Dirichlet boundary conditions associated to $S_{\cal K}$ 
(after some appropriate analytic continuation). 
Equivalently, we consider $M$ branes wrapping $S_{\cal K}$, where $M$ is the 
rank of $V$, and compute the partition function of topological string theory in this 
D-brane background. Since open string amplitudes have the BPS structure explained in \conj\ and \fr, 
this leads to structure results for the knot invariants $W_R^{U(N)}({\cal K})$ (which play the 
r\^ole of $Z_R$). This is explained in detail in \refs{\LMi,\LMV,\LMii}. 

The large $N$ duality of \GV\ can be generalized by considering an orientifold of the 
two geometries involved in the geometric transition, namely the resolved and the deformed conifold \SV. 
The deformed conifold is defined by the equation $z_1 z_4 - z_2 z_3=\mu$ and 
it contains an ${\bf S}^3$. If we wrap $2N$ branes on the
three-sphere, the spacetime description of the
open topological string theory is Chern-Simons theory on ${\bf S}^3$
with gauge group $U(2N)$ and at level $k$ (the level is related to the open string
coupling constant). We
now consider the following involution of the geometry
\eqn\stinv{
I:(z_1, z_2, z_3, z_4) \rightarrow (\bar z_4, -\bar z_3, -\bar z_2, \bar z_1)}
that leaves the
${\bf S}^3$ invariant. The string field
theory for the resulting open strings is now
Chern-Simons theory with gauge group $SO(N)$ or $Sp(N)$, depending on the choice of
orientifold action on the gauge group. The orientifold action 
on the resolved conifold is given by \orres. It then follows from the results of 
\OV\ and the orientifold action considered in \SV\ that the Chern-Simons generating functional 
\eqn\cssogen{
Z_{SO/Sp}(V)=\sum_R W_R^{SO/Sp}({\cal K}) {\rm Tr}_R \, V, 
}
where $V$ is again a $U(M)$ matrix, is the total partition function for open strings propagating 
on the orientifold of the resolved conifold with $M$ branes wrapping $S_{\cal K}$. In 
particular, the logarithm of \cssogen\ will have the structure explained in \freen, where the 
oriented contribution is obtained by considering a covering geometry 
with both $S_{\cal K}$ and its image under the involution \orres, 
$I(S_{\cal K})$. We can then translate the structure results presented 
in section 2 into structure results for the coloured Kauffman polynomial. 

The main problem in making this translation precise is that, given an arbitrary knot ${\cal K}$, 
we lack a precise prescription to compute the contribution of the covering amplitude. 
The covering amplitude ${\cal C}_{R_1 R_2}$ is defined as the oriented amplitude in the covering 
geometry in the presence of two sets of branes wrapping $S_{\cal K}$ and $I(S_{\cal K})$, 
with representations $R_1$, $R_2$, respectively. If one of the representations is trivial, 
we recover the oriented amplitude in the presence of $S_{\cal K}$, therefore ${\cal C}_{R}=W_R^{U(N)}({\cal K})$. 
But in the general case it is not obvious how to determine ${\cal C}_{R_1 R_2}$. Although there are 
proposals for the geometry of the Lagrangian submanifolds $S_{\cal K}$ \refs{\LMV,\T}, a  
direct Gromov-Witten computation of the corresponding open string amplitudes seems to be very difficult. One 
possible way of determining ${\cal C}_{R_1 R_2}$ would be to translate it into a pure knot-theoretic computation 
in the context of Chern-Simons theory, but we haven't found a completely satisfactory solution 
to this problem. 

Although we don't know how to compute the covering amplitude for an arbitrary knot, we can 
still extract the $\widehat f_{R}^{c=1}$ amplitudes from the knowledge of $W_{R}^{SO(N)}({\cal K})$. 
This goes as follows. Let us define the rational functions $g_R(q,\lambda)$ through the 
following equation
\eqn\gR{
\log Z_{SO}(V)=\sum_R \sum_{d \, {\rm odd}}{1\over d}g_R (q^d, \lambda^d)
{\rm Tr}_R V^d,}
and define as well
\eqn\hgR{
\widehat g_R (q,\lambda)=\sum_{R R'} M^{-1}_{R R'} (q)g_R (q, \lambda).}
Clearly, since we are not substracting the covering piece in the l.h.s. of \gR, we cannot expect 
much structure for $\widehat g_R (q,\lambda)$. However, one has that
\eqn\frcone{
\widehat f_R^{c=1}(q, \lambda)={1\over 2} \bigl(\widehat g_R (q,\lambda^{1\over 2})-
(-1)^{\ell(R)}\widehat g_R (q,-\lambda^{1\over 2})\bigr).
}
This follows from parity considerations. The invariants $W_R^{U(N)}({\cal K})$ have powers of $\lambda^{1\over 2}$ of the form 
$\ell(R) + 2k$, while $W_R^{SO(N)}({\cal K})$ have powers of $\lambda^{1\over 2}$ both of the form $\ell(R) + 2k$ and $\ell(R) + 2k+1$. 
The first ones correspond to both oriented and $c=2$ contributions, while the last ones correspond to $c=1$ 
contributions. Also, the covering contribution ${\cal C}_{R_1 R_2}$ (being an oriented amplitude) contains only powers 
in $\lambda^{1\over 2}$ of the form $\ell(R_1) + \ell(R_2) + 2k$. It is now easy to see from the 
results in section 2 that $\widehat f_R^{c=1}$ does not involve at all the covering contributions, and can be 
computed solely from the $SO(N)$ invariants, precisely in the way specified by \frcone. We can then formulate 
the following conjecture concerning the structure of the coloured Kauffman polynomial:

{\bf Conjecture}. Let $\widehat g_R (q, \lambda)$ be defined in terms of the coloured Kauffman polynomial 
by \gR\ and \hgR. Then, we have that
\eqn\conj{
{1\over 2} \bigl(\widehat g_R (q,\lambda^{1\over 2})-
(-1)^{\ell(R)}\widehat g_R (q,-\lambda^{1\over 2})\bigr)=\sum_{g,\beta} N^{c=1}_{R,g,\beta} 
(q^{1\over 2} -q^{-{1\over 2}})^{2g} \lambda^\beta,
}
where $N^{c=1}_{R,g,\beta}$ are {\it integer} numbers. They are 
BPS invariants corresponding to unoriented open string amplitudes with one crosscap.    
 
In the case of $W_{\tableau{1}}^{SO(N)}(q,\lambda)$, which is the unnormalized Kauffman polynomial, we can be slightly 
more precise, since we know that ${\cal C}_{\tableau{1}}(q,\lambda)=W_{\tableau{1}}^{U(N)}(q,\lambda)$, which is 
the unnormalized HOMFLY polynomial. We then deduce that
\eqn\HOMK{
W_{\tableau{1}}^{SO(N)}(q,\lambda)-W_{\tableau{1}}^{U(N)}(q,\lambda) = \sum_{g,\beta} N^{c=1}_{\tableau{1},g,\beta} 
(q^{1\over 2} -q^{-{1\over 2}})^{2g} \lambda^\beta + 
\sum_{g,\beta} N^{c=2}_{\tableau{1},g,\beta} (q^{1\over 2} -q^{-{1\over 2}})^{2g+1} \lambda^\beta.}
On the other hand, it follows from integrality of the oriented amplitudes that
\eqn\expH{
W_{\tableau{1}}^{U(N)}(q,\lambda)=\sum_{g\ge 0} p^H_g (\lambda) (q^{1\over 2} -q^{-{1\over 2}})^{2g-1},}
where $p^H_g (\lambda)$ is an odd polynomial in $\lambda^{\pm {1\over 2}}$. It then follows that 
the structure of the unnormalized Kauffman polynomial is given by
\eqn\expH{
W_{\tableau{1}}^{SO(N)}(q,\lambda)=\sum_{b\ge 0} p^K_b (\lambda) (q^{1\over 2} -q^{-{1\over 2}})^{b-1},}
where $p^K_b (\lambda)$ is an odd (even) polynomial in $\lambda^{\pm {1\over 2}} $ for $b$ even (odd). Moreover, 
\eqn\equalHK{
p_0^K(\lambda)=p_0^H(\lambda).}
This structural prediction turns out to be a well-known 
result in the theory of the Kauffman polynomial, see for example \L, page 183. 
One can easily compute $N^{c=1,2}_{\tableau{1},g,\beta}$ for various knots by computing the 
corresponding Kauffman polynomial. For example, the results of \LP\ imply that 
\eqn\twovan{
N^{c=2}_{\tableau{1},g,\beta}=0}
for all torus knots.  
 
Let us now turn to checks of the conjecture above for different knots and higher representations. 
The simplest case is of course the unknot, but 
this case has been already checked in section 3 (indeed, in the case of the unknot we know even how 
to compute the covering amplitude for arbitrary representations). In order to test the conjecture, we 
have to compute the invariants $W_{R}^{SO(N)}({\cal K})$ for arbitrary $R$. 
A class of nontrivial knots where this is doable are torus knots. In the case of $U(N)$ invariants, this 
was done in \LMi\ by using the formalism of knot operators \LLLR\ and the results of \LMlink. 
For $SO(N)$, the formalism of knot operators was used in \LP\ to compute invariants in the fundamental 
representation, but this has not been generalized to higher representations. For torus knots of the form $(2,m)$, 
however, one can use the results of \rama\ to write down a formula for the invariants in any 
representation of any gauge group. The formula reads as follows:
\eqn\twominv{
{\cal W}_{R}^G({\cal K}_{(2,m)}) =\sum_{S \in R \otimes R} ({\rm dim}_q S) (c_S(R, R))^m
}
where
\eqn\coeff{
c_S(R_1, R_2)= \epsilon_{R_1 R_2}^S q^{{C_{R_1} + C_{R_2} \over 2} -{C_S\over 4}}
}
In this equation, $C_R$ is the quadratic 
Casimir 
\eqn\quadcas{
C_R=\kappa_R + \ell(R) (N+a),}
where $a$ is given in \avalues, and $\epsilon_{R_1 R_2}^S$ is a sign which counts whether $S$ appears 
symmetrically or antisimmetrically in the tensor product $R_1 \otimes R_2$. 
In case $S$ appears with no multiplicity, there is an explicit expression for this sign 
given by \naculich\
\eqn\sign{
\epsilon_{R_1 R_2}^S= (-1)^{\rho\cdot(\Lambda_1 + \Lambda_2 -\Lambda_S)},
}
where $\Lambda_1$, $\Lambda_2$, $\Lambda_S$ are the 
highest weights of to the representations 
$R_1$, $R_2$, $S$, respectively. 
Using \twominv\ one can easily compute the invariants of the 
$(2,m)$ torus knots in the $SO(N)$ case, and extract $g_R$ (hence $N_{R,g,\beta}^{c=1}$) for various 
representations. In all cases we have found agreement with the above 
conjecture. We now present some results for the BPS invariants for the simplest torus knot, the $(2,3)$ knot or trefoil knot, 
for representations up to three boxes. All the invariants that are not shown in the Tables are understood to be zero.

\vskip 7mm

{\vbox{\ninepoint{
$$
\vbox{\offinterlineskip\tabskip=0pt
\halign{\strut
\vrule#
&
&\hfil ~$#$ \vrule
&\hfil ~$#$
&\hfil ~$#$
&\hfil ~$#$
&\vrule #\cr
\noalign{\hrule}
&
&\beta=1
&2
&3
&
\cr
\noalign{\hrule}
&g=0
&3
&-3
&1
&
\cr
&1
&1
&-1
&0
&
\cr
}\hrule}$$}
\vskip - 7 mm
\centerline{{\bf Table 1:} BPS invariants $N_{\tableau{1},g,\beta}^{c=1}$ for the trefoil knot.}
\vskip7pt}
\noindent
\smallskip

{\vbox{\ninepoint{
$$
\vbox{\offinterlineskip\tabskip=0pt
\halign{\strut
\vrule#
&
&\hfil ~$#$ \vrule
&\hfil ~$#$
&\hfil ~$#$
&\hfil ~$#$
&\hfil ~$#$
&\hfil ~$#$
&\vrule #\cr
\noalign{\hrule}
&
&\beta=3/2
&5/2
&7/2
&9/2
&11/2
&
\cr
\noalign{\hrule}
&g=0
&8
&-39
&69
&-53
&15
&
\cr
&1
&6
&-61
&146
&-126
&35
&
\cr
&2
&1
&-37
&128
&-120
&28
&
\cr
&3
&0
&-10
&56
&-55
&9
&
\cr
&4
&0
&-1
&12
&-12
&1
&
\cr
&5
&0
&0
&1
&-1
&0
&
\cr
}\hrule}$$}
\vskip - 7 mm
\centerline{{\bf Table 2:} BPS invariants $N_{\tableau{2},g,\beta}^{c=1}$ for the trefoil knot.}
\vskip7pt}
\noindent
\smallskip

{\vbox{\ninepoint{
$$
\vbox{\offinterlineskip\tabskip=0pt
\halign{\strut
\vrule#
&
&\hfil ~$#$ \vrule
&\hfil ~$#$
&\hfil ~$#$
&\hfil ~$#$
&\hfil ~$#$
&\hfil ~$#$
&\vrule #\cr
\noalign{\hrule}
&
&\beta=3/2
&5/2
&7/2
&9/2
&11/2
&
\cr
\noalign{\hrule}
&g=0
&16
&-69
&111
&-79
&21
&
\cr
&1
&20
&-146
&307
&-251
&70
&
\cr
&2
&8
&-128
&366
&-330
&84
&
\cr
&3
&1
&-56
&230
&-220
&45
&
\cr
&4
&0
&-12
&79
&-78
&11
&
\cr
&5
&0
&-1
&14
&-14
&1
&
\cr
&6
&0
&0
&1
&-1
&0
&
\cr
}\hrule}$$}
\vskip - 7 mm
\centerline{{\bf Table 3:} BPS invariants $N_{\tableau{1 1},g,\beta}^{c=1}$ for the trefoil knot.}
\vskip7pt}
\noindent
\smallskip

The results for representations with three boxes are listed in Appendix C. 

Although we have focused in this section on the case of knots, it is straightforward 
to extend the conjecture above to the case of links, and extract the $c=1$ piece from the 
$SO$ Chern-Simons invariants. Framed knots can be also considered by using exactly the 
same rules that are used for $U(N)$ invariants \MV.

\newsec{Discussion}

In this paper we extended the results of \BFM\ in order to study open string amplitudes on 
orientifolds without fixed planes. We found the general structure of the twisted and untwisted 
contributions, we determined the BPS structure of the corresponding amplitudes, and we checked 
our results in various examples. We want to remark that, although our main testing ground 
have been orientifolds of noncompact, toric Calabi-Yau orientifolds with noncompact 
branes, the general results about the structure and integrality properties of 
the amplitudes should be valid in general. 

One of the motivations of the present paper was to extend the results of \refs{\LMV,\LMii} on the 
BPS structure of the coloured HOMFLY polynomial to the coloured Kauffman polynomial. Although our general 
structural results on open string amplitudes on orientifolds give a first principles answer to this problem, 
as it has been made clear in the analysis of the framed unknot and the Hopf link,  
we haven't been able to determine the covering contribution for arbitrary knots. This is an important 
open issue that one should resolve in order to obtain a complete picture of the correspondence 
between enumerative geometry and knot invariants implied by large $N$ dualities.  

We have also seen that the predictions obtained from the topological vertex in the unoriented case 
agree with unoriented localization computations in the examples that we have worked out. It would 
be very interesting to derive a more general and precise correspondence between these two approaches, 
following the lines of the mathematical treatment of the vertex given in \refs{\DF,\LLLZ}, and maybe 
connect the unoriented Gromov-Witten theory sketched here and in \refs{\DFM,\BFM}, 
with a moduli problem involving ideal sheaves, generalizing in this way the results of \MNOP. 
Finally, our results both in this paper and in \BFM\ cover only orientifolds without fixed planes, and 
more work is needed in order to understand the general situation from the point of view of topological string 
theory.

\bigskip
\centerline{\bf Acknowledgments }\nobreak
\bigskip

M.M. would like to thank Cumrun Vafa and specially Jos\'e Labastida for discussions on 
$SO(N)$ Chern-Simons invariants and topological strings. The work of V.B. was supported by a Rhodes Scholarship and a NSERC PGS Doctoral Scholarship. The work of B.F. was supported in part by DOE grant DE-FG02-96ER40959.

\appendix{A}{Useful formulae and Schur functions.}

In this appendix we will list some useful identities of Schur functions and their relations to the unknot and Hopf link invariants. For a more 
detailed discussion of Schur functions see for example \refs{\Mac,\FH}. Applications of these results to topological 
string computations can be found in \refs{\ORV,\EK,\HIV,\IKStrip}.

Let $R$ be a partition associated to a Young tableau. Let $\ell(R)$ be the number of boxes of the Young tableau and $l_i (R)$ be the number of boxes in the $i$-th row. We define the quantity
\eqn\unknotschur{
W_R (q) = s_R (q^\rho)
}
where $s_R (q^\rho)$ is the Schur function with the substitution $s_R( x_i = q^{-i + 1/2})$, where $i$ runs from $1$ to $\infty$. 
$W_R (q)$ is the leading order of the $U(N)$ quantum dimension $\dim_q^{U(N)} R$ (in the sense defined in \AMV). We also recall the general formula 
for quantum dimensions of a group $G$:
 \eqn\qdgeneral{
\dim_q^G R = \prod_{\alpha \in \Delta_+} {[(\Lambda_R + \rho, \alpha)] \over [(\rho, \alpha)]},
}
where $\Lambda_R$ is the highest weight of the representation $R$, $\rho$ is the Weyl vector, and the 
product is over the positive roots of $G$ . We also defined the
following $q$-number:
\eqn\qnumber{
[x]=q^{x/2} - q^{-x/2}.
}
Another important object is 
\eqn\hopfschur{
W_{R_1 R_2} (q) = s_{R_1} ( q^{\rho}) s_{R_2} (q^{\ell(R_1) + \rho} ),
}
where $s_{R_2} (q^{\ell(R_2) + \rho}) = s_{R_2} (x_i = q^{l_i (R_2)-i+1/2})$. This is the 
leading part (again in the sense of \AMV) of the Hopf link invariant ${\cal W}_{R_1 R_2}^{U(N)}$.

The topological vertex formula derived in \AKMV\ reads
\eqn\topvertex{
C_{R_1 R_2 R_3} = q^{{1 \over 2}(\kappa_{R_2}+\kappa_{R_3})} \sum_{Q_1,Q_2, R} N_{Q_1 R}^{R_1} N_{Q_2 R}^{R_3^t} 
{W_{R_2^t Q_1} W_{R_2 Q_2} \over W_{R_2}},
}
where $\kappa_R$ is defined by $\kappa_R=\sum_i l_i(R) (l_i(R) -2i +1)$. Using \unknotschur\ and \hopfschur\ we can express the topological vertex in terms of Schur functions (this was first done in \ORV)
\eqn\topvertexschur{
C_{R_1 R_2 R_3} = q^{{1 \over 2}(\kappa_{R_2}+\kappa_{R_3})} s_{R_2^t} (q^\rho) \sum_{Q} s_{R_1/Q} (q^{\ell(R_2^t) + \rho}) s_{R_3^t / Q} (q^{\ell(R_2) + \rho}),
}
where we have used skew Schur functions defined as
\eqn\skewschur{
s_{R/R_1} (x)= \sum_{Q} N_{R_1 Q}^R s_Q (x).
}

Schur functions satisfy some useful identities. First, we have
\eqn\identityone{
s_{R^t} (q) = q^{-\kappa_R/2} s_{R} (q) = (-1)^{\ell (R)} s_R (q),
}
and similarly
\eqn\identitytwo{
s_{R/R_1} (q)= (-1)^{\ell(R)-\ell(R_1)} s_{R^t/R_1^t} (q).
}
The two following formulae are also important:
\eqn\identitythree{\eqalign{
\sum_R s_{R/R_1}(x) s_{R/R_2} (y) &= \prod_{i,j \geq 1} (1-x_i y_j)^{-1} \sum_Q s_{R_2/Q} (x) s_{R_1/Q} (y),\cr
\sum_R s_{R/R_1}(x) s_{R^t/R_2} (y) &= \prod_{i,j \geq 1} (1+x_i y_j) \sum_Q s_{R_2^t/Q} (x) s_{R_1^t/Q^t} (y).\cr
}}
The following result was proved in \EK. Let us define the ``relative" hook length
\eqn\hook{
h_{R_1 R_2} (i,j) = l_i (R_1) +l_j (R_2) -i -j +1,
}
and the following functions
\eqn\fctschur{\eqalign{
f_R(q) &= {q \over (q-1)} \sum_{i \geq 1} (q^{l_i(R)-i} - q^{-i}),\cr
{\tilde f}_{R_1 R_2} (q) &= {(q-1)^2\over q} f_{R_1} (q) f_{R_2} (q) +f_{R_1}(q) + f_{R_2} (q).
}}
Let us denote the expansion coefficients of ${\tilde f}_{R_1 R_2} (q)$ by
\eqn\expcoeff{
{\tilde f}_{R_1 R_2} (q) = \sum_k C_k (R_1, R_2) q^k.
}
Alternatively, 
\eqn\coeffs{
\sum_k C_k (R_1, R_2) q^k={ W_{R_1 \tableau{1}} \over W_{R_1}}
{ W_{R_2 \tableau{1}} \over W_{R_2}}- W_{\tableau{1}}^2.}
Then it was proved that
\eqn\identityfour{
\prod_{i,j \geq 1} (1-Q q^{h_{R_1 R_2} (i,j)} ) = \prod_{k=1}^{\infty} (1-Q q^k)^k \prod_k (1-Q q^k)^{C_k (R_1, R_2)}.
}

Let us now present a useful result proved by Littlewood \refs{\Li, \Mac}:
\eqn\littlewood{
\sum_{R=R^T} s_R (x) (-1)^{{1\over2} (\ell(R)\mp r(R))} = \prod_{i=1}^{\infty} (1 \pm x_i) \prod_{1 \leq i < j < \infty} (1-x_i x_j),
}
where $r(R)$ is the rank of $R$. The final formula that we will need reads as follows \HIV\
\eqn\exponential{
\prod_{i,j} (1-Q x_i y_j) = \exp \Big{[}- \sum_{n=1}^{\infty} {Q^n \over n} \sum_{i,j} x_i^n y_j^n\Big{]},
}
from which we can deduce the identities
\eqn\expident{\eqalign{
\prod_i (1\mp Q^{1/2} q^{i-1/2}) =& \exp \Big{[} \sum_{n=1}^{\infty} {(\pm 1)^n Q^{n/2} \over n (q^{n/2}-q^{-n/2})} \Big{]},\cr
\prod_{i,j} (1-Q q^{i+j-1}) =& \exp \Big{[}- \sum_{n=1}^{\infty} {Q^{n} \over n (q^{n/2}-q^{-n/2})^2} \Big{]}.
}}

\appendix{B}{A useful identity}

In our previous paper \BFM\ we conjectured the relation \magic\ between the topological vertex and $SO$/$Sp$ Chern-Simons expectation values 
of Hopf links. This identity showed that the new vertex rule introduced in \BFM\ to compute amplitudes on orientifolds 
agrees with the results of large $N$ $SO/Sp$ transitions. In \BFM\ we only presented a partial proof of this relation; we will now present the 
full proof. We will only consider here the $Sp$ case for the sake of clarity, but the proof for the $SO$ case is similar.

Let us start by considering \magic\ for trivial representations $R_1=R_2=\cdot$. In this case we have to show that
\eqn\triviali{
\sum_{R=R^T} C_{R \cdot \cdot} Q^{\ell(R)/2} (-1)^{{1\over 2}(\ell(R)+r(R))}=S_{00}^{{\rm Sp(N)}}.
}
The l.h.s. can be rewritten using \topvertexschur, \identityone\ and \littlewood\ as
\eqn\trivialii{\eqalign{
\sum_{R=R^T} s(Q^{1/2}q^{-\rho}) (-1)^{{1\over 2}(\ell(R)-r(R))}&= \prod_{i=1}^{\infty} (1+Q^{1/2} q^{i-1/2}) \prod_{1\leq i < j < \infty} (1-Q q^{i+j-1})\cr
&= {\prod_{i=1}^{\infty} (1+Q^{1/2} q^{i-1/2})^{1/2}  \prod_{i,j,=1}^{\infty} (1-Q q^{i+j-1})^{1/2}\over \prod_{i=1}^{\infty} (1-Q^{1/2} q^{i-1/2})^{1/2}} 
}}
Using \expident\ we find that the r.h.s. becomes
\eqn\trivialiii{
\exp \Big{[}{1 \over 2} \sum_{n=1}^{\infty} {((-1)^n -1) Q^{n/2} \over n (q^{n/2}-q^{-n/2})} \Big{]} \exp \Big{[}-{1 \over 2} \sum_{n=1}^{\infty} {Q^{n} \over n (q^{n/2}-q^{-n/2})^2} \Big{]}
}
which is exactly $S_{00}^{\rm{Sp(N)}}$, so \triviali\ is proved.

Let now $R_1=\cdot$ be the trivial representation and $R_2=\mu$ be any representation. We must now show that
\eqn\trivonei{
{1 \over S_{00}^{{\rm Sp(N)}}} \sum_{R=R^T} C_{\cdot \mu^T R} Q^{{1 \over 2} (\ell(R)-\ell(\mu))} q^{\kappa_{\mu} \over 2} (-1)^{{1\over 2}(\ell(R)+r(R))}= \CW_{\mu}^{\rm Sp(N)}.
}
Using \topvertexschur\ and \identityone\ the l.h.s can be rewritten as
\eqn\trivoneii{
{1 \over S_{00}^{{\rm Sp(N)}}} s_{\mu} (Q^{-1/2} q^{\rho}) \sum_{R=R^T} (-1)^{{1\over 2}(\ell(R)-r(R))} s_R (Q^{1/2} q^{-\ell(\mu)-\rho}).
}
From \littlewood, the first line of \trivialii\ and the definition of $W_R (q) = s_R (q^\rho)$ in terms of q-numbers (see eq. (7.5) of \AKMV), we find, after some algebra:
\eqn\trivoneiii{\eqalign{
&Q^{-\ell(\mu)/2} \prod_{1\leq i< j \leq d(\mu)} {[l_i+l_j-i-j+1]_{Q^{-1}} [l_i-l_j+j-i] \over [-i-j+1]_{Q^{-1}} [j-i]} \cr
&\times \prod_{i=1}^{d(\mu)} {[1-i]_{Q^{-1}}^{\rm Sp(N)} [2l_i-2i+1]_{Q^{-1}} \over [l_i + 1-i]_{Q^{-1}}^{\rm Sp(N)} [-2i+1]_{Q^{-1}}} \prod_{v=1}^{l_i} Q^{1/2} {[l_i-i-v-d(\mu)+1]_{Q^{-1}} \over [v-i+d(\mu)]}
}}
where $d(\mu)$ is the number of rows of $\mu$, and we used the q-numbers $[x] = q^{x/2}-q^{-x/2}$, $[x]_{\lambda} = \lambda^{1/2} q^{x/2}-\lambda^{-1/2} q^{-x/2}$ and $[x]_{\lambda}^{\rm Sp(N)} = \lambda^{1/4} q^{{1\over 4}(2x-1)} - \lambda^{-1/4} q^{-{1 \over 4}(2x-1)}$. One can see that the two factors of $Q$ cancel out of \trivoneiii, and the remaining expression is exactly the definition of the $Sp(N)$ quantum dimension of $\mu$ for $\lambda = Q^{-1}$ (see eq. (4.9) in \BFM). But $\CW_{\mu}^{\rm Sp(N)} = \dim_q^{\rm Sp(N)} \mu (\lambda=Q^{-1})$. Therefore \trivonei\ is proved.

We are now in position to prove \magic\ in the general case, namely we have to show that
\eqn\magicii{
{1\over S_{00}^{Sp(N)}}\sum_{R=R^T}   C_{R_1 R_2^T R} Q^{\ell(R)/2}
(-1)^{{1\over 2}(\ell(R) + r(R))} = q^{-{\kappa_{R_2}\over 2}}
Q^{{1\over 2}(\ell(R_1) + \ell (R_2))}
{\cal W}^{Sp(N)}_{R_1 R_2}.}
Let us first rewrite the Hopf link expectation value in terms of quantum dimensions, using eq. (4.19) of \BFM. The r.h.s. becomes:
\eqn\hopfquant{
\sum_{\mu,\lambda_1,\lambda_2,\lambda_3} N^{R_1}_{\lambda_1 \lambda_2} N^{R_2}_{\lambda_1 \lambda_3} N^{\mu}_{\lambda_2 \lambda_3} q^{{1\over2}(\kappa_{R_1}-\kappa_{\mu})}
Q^{{1\over 2}(\ell(\mu))} {\cal W}^{Sp(N)}_{\mu},
}
where we expressed the $Sp$ tensor product coefficients in terms of Littlewood-Richardson coefficients \FH. We can now rewrite the r.h.s. using \trivonei\ and \topvertexschur\ as 
\eqn\rhsschur{\eqalign{
{1\over S_{00}^{Sp(N)}}& \sum_{R=R^T} Q^{{1\over2}\ell(R)} (-1)^{{1\over 2}(\ell(R)+ r(R))}s_{R^T} (q^{\rho})  q^{{1\over2}(\kappa_{R_1}+\kappa_{R})}\cr
&\times \sum_{\mu,\lambda_1,\lambda_2,\lambda_3} N^{R_1}_{\lambda_1 \lambda_2} N^{R_2}_{\lambda_1 \lambda_3} N^{\mu}_{\lambda_2 \lambda_3} s_{\mu^T}(q^{\ell(R)+\rho})
}}
The sum in the second line can be explicitely evaluated by using \identityone, \identitytwo, the definition of skew Schur functions \skewschur\ and the fact that $s_{R_1} (x) s_{R_2} (x) = \sum_{R} N^{R}_{R_1 R_2} s_R (x)$:
\eqn\sumeval{
\sum_{\mu,\lambda_1,\lambda_2,\lambda_3} N^{R_1}_{\lambda_1 \lambda_2} N^{R_2}_{\lambda_1 \lambda_3} N^{\mu}_{\lambda_2 \lambda_3} s_{\mu^T}(q^{\ell(R)+\rho})
 =\sum_{\lambda_1} s_{R_1^T/\lambda_1^T} (q^{\ell(R)+\rho}) s_{R_2^T/\lambda_1^T} (q^{\ell(R) + \rho})
}
 Inserting \sumeval\ in \rhsschur\ gives (using the fact that $R=R^T$):
\eqn\rhsfinal{\eqalign{
{1\over S_{00}^{Sp(N)}} & \sum_{R=R^T} Q^{{1\over2}\ell(R)} (-1)^{{1\over 2}(\ell(R) + r(R))} \cr
 &\times {\left[ q^{{1\over2}(\kappa_{R_1}+\kappa_{R})}s_{R^T} (q^{\rho}) \sum_{\lambda_1}s_{R_2^T/\lambda_1} (q^{\ell(R^T) + \rho}) s_{R_1^T/\lambda_1} (q^{\ell(R)+\rho})  \right]}
}}
The term in brackets is exactly the definition of $C_{R_2^T R R_1}$ in terms of Schur functions (see \topvertexschur). Therefore \rhsfinal\ is equal to the l.h.s. of \magicii\ and the identity \magic\ is proved.

\appendix{C}{BPS invariants for the trefoil knot}
In this Appendix, we list the BPS invariants $N^{c=1}_{R,g,\beta}$ for representations $R$ with three boxes.

{\vbox{\ninepoint{
$$
\vbox{\offinterlineskip\tabskip=0pt
\halign{\strut
\vrule#
&
&\hfil ~$#$ \vrule
&\hfil ~$#$
&\hfil ~$#$
&\hfil ~$#$
&\hfil ~$#$
&\hfil ~$#$
&\hfil ~$#$
&\hfil ~$#$
&\hfil ~$#$
&\vrule #\cr
\noalign{\hrule}
&
&\beta=2
&3
&4
&5
&6
&7
&8
&9
&
\cr
\noalign{\hrule}
&g=0
&18
&-270
&1185
&-2380
&2430
&-1188
&175
&30
&
\cr
&1
&21
&-753
&4924
&-12209
&13203
&-4856
&-1300
&970
&
\cr
&2
&8
&-1007
&10374
&-31348
&31419
&4028
&-22155
&8681
&
\cr
&3
&1
&-793
&13920
&-50383
&30636
&84956
&-117415
&39078
&
\cr
&4
&0
&-378
&12688
&-54222
&-24584
&305272
&-343318
&104542
&
\cr
&5
&0
&-106
&8006
&-40151
&-118255
&609701
&-639896
&180701
&
\cr
&6
&0
&-16
&3486
&-20657
&-178503
&797521
&-813994
&212163
&
\cr
&7
&0
&-1
&1024
&-7353
&-161931
&728309
&-734484
&174436
&
\cr
&8
&0
&0
&193
&-1773
&-98947
&478948
&-480509
&102088
&
\cr
&9
&0
&0
&21
&-276
&-42205
&229955
&-230209
&42714
&
\cr
&10
&0
&0
&1
&-25
&-12624
&80705
&-80729
&12672
&
\cr
&11
&0
&0
&0
&-1
&-2599
&20474
&-20475
&2601
&
\cr
&12
&0
&0
&0
&0
&-351
&3654
&-3654
&351
&
\cr
&13
&0
&0
&0
&0
&-28
&435
&-435
&28
&
\cr
&14
&0
&0
&0
&0
&-1
&31
&-31
&1
&
\cr
&15
&0
&0
&0
&0
&0
&1
&-1
&0
&
\cr
}\hrule}$$}
\vskip - 7 mm
\centerline{{\bf Table 4:} BPS invariants $N_{\tableau{3},g,\beta}^{c=1}$ for the trefoil knot.}
\vskip 7pt}
\noindent
\smallskip

{\vbox{\ninepoint{
$$
\vbox{\offinterlineskip\tabskip=0pt
\halign{\strut
\vrule#
&
&\hfil ~$#$ \vrule
&\hfil ~$#$
&\hfil ~$#$
&\hfil ~$#$
&\hfil ~$#$
&\hfil ~$#$
&\hfil ~$#$
&\hfil ~$#$
&\hfil ~$#$
&\vrule #\cr
\noalign{\hrule}
&
&\beta=2
&3
&4
&5
&6
&7
&8
&9
&
\cr
\noalign{\hrule}
&g=0
&99
&-1125
&4359
&-8096
&7828
&-3699
&563
&72
&
\cr
&1
&201
&-4194
&22748
&-51475
&53807
&-21649
&-2204
&2766
&
\cr
&2
&164
&-7702
&60811
&-165827
&171590
&-19997
&-68978
&29939
&
\cr
&3
&66
&-8701
&104757
&-338906
&282625
&264688
&-468878
&164349
&
\cr
&4
&13
&-6395
&125047
&-472907
&124226
&1398430
&-1710505
&542091
&
\cr
&5
&1
&-3092
&106648
&-466523
&-477321
&3645201
&-3976290
&1171376
&
\cr
&6
&0
&-971
&65795
&-331606
&-1232410
&6113672
&-6363573
&1749093
&
\cr
&7
&0
&-190
&29358
&-171307
&-1590490
&7192295
&-7328205
&1868539
&
\cr
&8
&0
&-21
&9358
&-64261
&-1351903
&6186865
&-6240225
&1460187
&
\cr
&9
&0
&-1
&2072
&-17298
&-815116
&3979137
&-3994110
&845316
&
\cr
&10
&0
&0
&302
&-3252
&-358192
&1934294
&-1937220
&364068
&
\cr
&11
&0
&0
&26
&-405
&-115397
&712126
&-712504
&116154
&
\cr
&12
&0
&0
&1
&-30
&-26996
&197286
&-197315
&27054
&
\cr
&13
&0
&0
&0
&-1
&-4465
&40454
&-40455
&4467
&
\cr
&14
&0
&0
&0
&0
&-495
&5952
&-5952
&495
&
\cr
&15
&0
&0
&0
&0
&-33
&594
&-594
&33
&
\cr
&16
&0
&0
&0
&0
&-1
&36
&-36
&1
&
\cr
&17
&0
&0
&0
&0
&0
&1
&-1
&0
&
\cr
}\hrule}$$}
\vskip - 7 mm
\centerline{{\bf Table 5:} BPS invariants $N_{\tableau{2 1},g,\beta}^{c=1}$ for the trefoil knot.}
\vskip7pt}
\noindent
\smallskip

{\vbox{\ninepoint{
$$
\vbox{\offinterlineskip\tabskip=0pt
\halign{\strut
\vrule#
&
&\hfil ~$#$ \vrule
&\hfil ~$#$
&\hfil ~$#$
&\hfil ~$#$
&\hfil ~$#$
&\hfil ~$#$
&\hfil ~$#$
&\hfil ~$#$
&\hfil ~$#$
&\vrule #\cr
\noalign{\hrule}
&
&\beta=2
&3
&4
&5
&6
&7
&8
&9
&
\cr
\noalign{\hrule}
&g=0
&108
&-1044
&3705
&-6484
&6000
&-2754
&427
&42
&
\cr
&1
&306
&-4818
&23074
&-48785
&49436
&-20669
&-448
&1904
&
\cr
&2
&366
&-11012
&73663
&-186538
&193691
&-44683
&-49616
&24129
&
\cr
&3
&230
&-15636
&151596
&-453623
&421630
&161750
&-421269
&155322
&
\cr
&4
&79
&-14720
&216949
&-756616
&429479
&1359478
&-1836601
&601952
&
\cr
&5
&14
&-9381
&223615
&-898781
&-235791
&4434624
&-5047078
&1532778
&
\cr
&6
&1
&-4047
&168943
&-777340
&-1531480
&8961515
&-9525899
&2708307
&
\cr
&7
&0
&-1160
&94128
&-495542
&-2661004
&12577678
&-12957296
&3443196
&
\cr
&8
&0
&-211
&38523
&-233794
&-2843448
&12900213
&-13087921
&3226638
&
\cr
&9
&0
&-22
&11409
&-81283
&-2124814
&9936047
&-10004126
&2262789
&
\cr
&10
&0
&-1
&2373
&-20525
&-1160684
&5832726
&-5850601
&1196712
&
\cr
&11
&0
&0
&328
&-3656
&-470990
&2625946
&-2629249
&477621
&
\cr
&12
&0
&0
&27
&-435
&-142042
&905758
&-906165
&142857
&
\cr
&13
&0
&0
&1
&-31
&-31433
&237305
&-237335
&31493
&
\cr
&14
&0
&0
&0
&-1
&-4959
&46375
&-46376
&4961
&
\cr
&15
&0
&0
&0
&0
&-528
&6545
&-6545
&528
&
\cr
&16
&0
&0
&0
&0
&-34
&630
&-630
&34
&
\cr
&17
&0
&0
&0
&0
&-1
&37
&-37
&1
&
\cr
&18
&0
&0
&0
&0
&0
&1
&-1
&0
&
\cr
}\hrule}$$}
\vskip - 7 mm
\centerline{{\bf Table 6:} BPS invariants $N_{\tableau{1 1 1},g,\beta}^{c=1}$ for the trefoil knot.}
\vskip7pt}
\noindent
\smallskip

\listrefs

\bye